# Why orb-weaving spiders use leg crouching behavior in vibration sensing of prey on a web: A physical mechanism from robophysical modeling

Eugene H. Lin[1*], Yishun Zhou[1], Hsin-Yi Hung[2], Luke Moon[1], Andrew Gordus[2,3], Chen Li[1*]

[1] Department of Mechanical Engineering, Johns Hopkins University, Baltimore, MD 21218, USA

[2] Solomon H. Snyder Department of Neuroscience, Johns Hopkins University, Baltimore, MD, 21205, USA

[3] Department of Biology, Johns Hopkins University, Baltimore, MD 21218, USA

**ABSTRACT.**

One of the key functions of organisms is to sense their physical environment so that they can react upon the sensed information appropriately. All spiders can perceive their environment through vibration sensors in their legs, and most spiders rely on substrate-born vibration sensing to detect prey. Orb-weaving spiders primarily sense leg vibrations to detect and locate prey caught on their wheel-shaped webs. Biological experiments and computational modeling elucidated the physics of how these spiders use long-timescale web-building behaviors, which occur *before* prey capture, to modulate vibration sensing of prey by controlling web geometry, materials, and tension distribution. By contrast, the physics of how spiders use short-timescale leg behaviors to modulate vibration sensing on a web *during* prey capture is less known. This is in part due to challenges in biological experiments (e.g., having little control over spider behavior, difficulty measuring the whole spider–web–prey system vibrations) and theoretical/computation modeling (e.g., close-form equations intractable for a complex web, high computation cost for simulating vibrations with behaving animals). Here, we use robophysical modeling as a complementary approach to address these challenges and study how dynamic leg crouching behavior common in orb-weaving spiders contributes to vibration sensing of prey on a web. Following observations in the orb-weaver *Uloborus diversus* from a parallel biological study, we created a robophysical model consisting of a spider robot that can dynamically crouch its legs and sense its leg vibrations and a prey robot that can shake both on a horizontal physical wheel-shaped web. Without the prey robot, after each dynamic crouch, the spider robot sensed leg vibrations with only one dominant frequency—the natural frequency of itself passively vibrating on the web. With the prey robot, after each dynamic crouch, the spider robot sensed leg vibrations with two dominant frequencies—the additional higher frequency being the natural frequency of itself passively vibrating on its spiral thread induced by the spider robot's dynamic crouch. This additional frequency increased as the prey robot became closer from the web center where the spider robot was. These features allowed the spider robot to detect prey presence and distance. We developed a minimalistic physics model that decoupled the spider–web–prey system into two subsystems to explain these observations. Guided by both these results, we found evidence of the same physical mechanism appearing in the web of the *U. diversus* spider during prey capture in the data from the parallel biological study. Our work demonstrated that robophysical modeling is a useful approach for discovering physical mechanisms of how spiders use short-time scale leg behaviors to enhance vibration sensing of objects on a web and providing new biological hypotheses.

I. INTRODUCTION.

In 1879, a curious physicist wondered what would happen if he held a tuning fork at a garden spider's web, while observing them spinning their webs [1]. He found that spiders react to the vibrating tuning fork in the same manner as to a ensnared insect, locating the fork and treating it like prey [1]. Since then, many biologists have investigated how spiders use and sense vibrations, with much progress on the sensory biology, neurophysiology, behavioral descriptions, and hypothesized physical mechanisms [2–4].

Vibration sensing is often the primary sensory modality for spiders to detect prey [5]. While some families can use vision (e.g., ogre-faced spiders [6,7], jumping spiders [8]), sound (e.g., jumping spiders [9], orb-weaver spiders [10]), air flow sensing (e.g., wandering spiders [11]), or chemical senses (e.g., wolf spiders [12]), most spiders heavily rely on sensing substrate-borne vibrations to detect prey [13]. For substrate-borne vibration sensing, spiders detect minute vibrations through sensory organs at their distal leg joints, known as lyriform organs. These consist of many slit sensilla near leg joints that can detect minute strains [3,14]. While most research on the sensory organs have focused on *Cupiennius salei* spiders that use vibration sensing to hunt on banana plants [3], their findings also apply to other spiders [13,15].

Many spiders use webs to extend their vibration sensing [16,17]. All known spiders species (53544 at the time of writing [18]) use silk to some extent [14], and about half of them use them to make webs for prey capture [19]. Spiders make many forms of webs for prey capture, including funnel webs, sheet webs, tangle webs, and orb-webs [4]. In this paper, to elucidate vibration-sensing in webs, we focus on the behavior of orb-weaving spiders (4700+ species [18]) that make the commonly known webs shaped like a wheel.

Orb-weaving spiders are functionally blind and rely on sensing vibrations to detect changes in their webs to sense their environments [20–27]. Using vibration sensing, they are extremely quick and proficient at detecting, localizing, and approaching prey on their web [27]. They can also use vibration sensing to identify other types of objects on their web, distinguishing between unwanted detritus [28], conspecific spiders as potential mate [29–32], offsprings [33], unwanted animals [33], and even invading predators [28]. Even non-orb-weaving spiders can use vibration sensing of webs to their own benefit, such as identifying orb-weavers' prey-wrapping behavior to steal prey off their webs [34], mating communication [35–37], or mimicking prey to bait orb-weavers to the periphery of their webs to eat them [38,39]. They can also use web vibrations to detect sounds around the web [10], although here we will focus on the web vibrations caused by prey caught in the web.

By using their webs to extend their sensory system [16], web-weaving spiders can understand their surroundings by the vibrations transmitted in the web [2,22,24,27]. Different vibration sources such as wind, prey, mate, offspring, and inanimate objects on the web produce different frequency components, and it is thought that the spider can use this information to distinguish them [27,32]. During prey capture, web-weaving spiders can use many strategies to modulate their sensory system extended through their webs to perceive their environment. Broadly, these fall into two categories: strategies involving long-timescale (hours [2,40]) web-building behaviors, and those involving short-timescale (seconds) leg behaviors during sensing and capture of prey caught on the web.

During long-timescale web building behaviors, web-weaving spiders can modulate the vibration dynamics of the web by changing the web's geometry, viscoelastic properties (via using different types of silk), and tension distribution [2]. Quantification of web vibration dynamics via experiments and finite element analysis modeling found that different web properties can influence the propagation of vibration waves in the web to identify and locate prey [17,22,23,41].

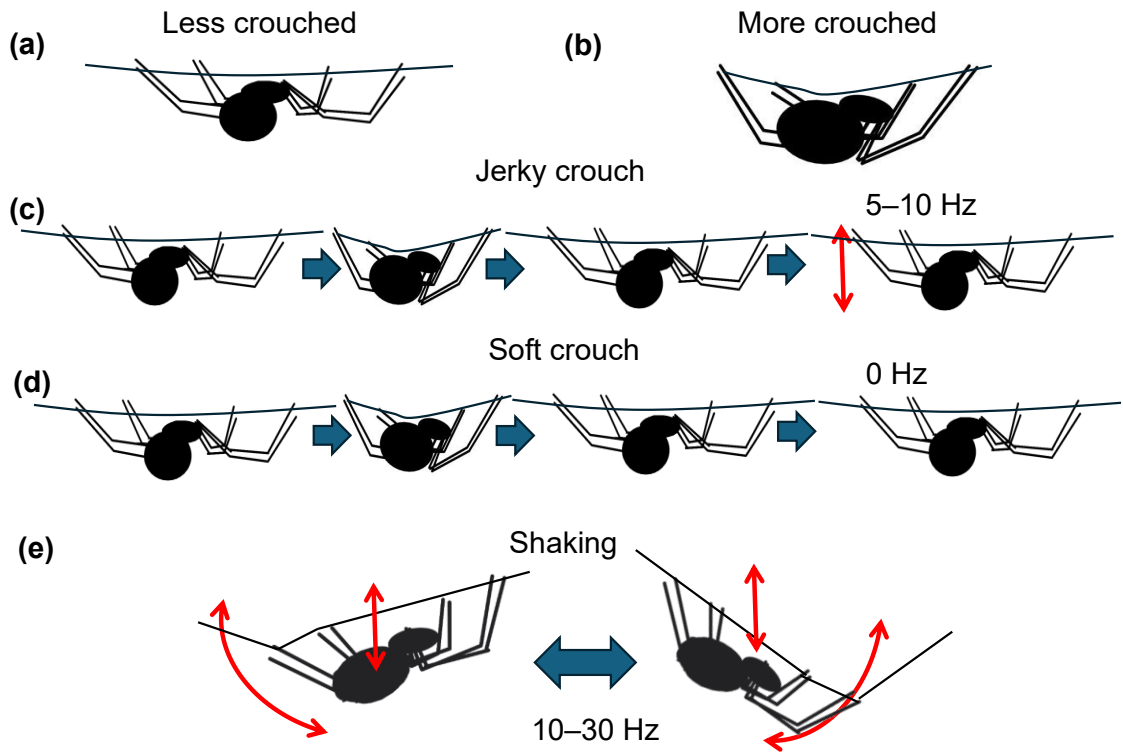

**Figure 1. Classification of short-timescale leg behaviors of spider vibration sensing. (a, b)** Assuming different static leg postures, either less (a) or more (b) crouched. **(c, d)** Dynamically crouching the legs,
44

either using a jerky crouch (c), with a subsequent passive vibration (5–10 Hz for *Uloborus diversus*), or a soft crouch (d), with no subsequent passive vibrations (0 Hz). **(e)** Using legs to shake the web.

Once a web is built, the web-weaving spider responds to prey caught on the web with short-timescale leg behaviors. Besides using legs to locomote on the web to reorient and move towards or from the other animal and to wrap prey with silk, the spider may also use leg behaviors not for locomotion to enhance vibration sensing on the web. Here, for providing context of our study, we categorize the latter leg behaviors for prey capture as follows:

1. Assuming different static postures (Fig. 1(a)). Some spiders rest in numerous postures on the web, such as a crouched posture when hungry [42]. This has also been observed in non-prey capture treatments, such as a lowered-abdomen posture during courtship, and a folded leg posture when the web is disturbed [42,43]. Adjusting postures not only changes inertial properties but may also change leg joint stiffness and damping (through different muscle activation in different postures) [44,45]. Both these changes may turn the spider's body and legs into a reconfigurable mechanical filter, which may allow it to "tune in" to specific vibration frequencies [42,46].

2. Dynamic crouching (Fig. 1(b) and (c)). During prey capture [25,26], the spider dynamically crouches its legs so that its abdomen moves closer to the web, then immediately returns to the initial, not crouched posture (recovery phase), without shaking the abdomen at a high frequency. Similar behaviors have also been observed during sexual signaling [32,47] or when the spider is identifying artificial stimuli [48]. The spider can do this repeatedly with intermittent pauses in between each crouch. Dynamic crouching has two sub-types with differences in the recovery phase.

    a. Jerky crouch (Fig. 1(b), Movie S1 [165]). The spider can release its legs quickly so that the web vibrates afterwards [47,49–51], causing the prey to also passively vibrate [50].

    b. Soft crouch (Fig. 1(c), Movie S2 [165]). The spider can control its legs during recovery so that the web vibrates little after the spider stops [50,52].

It is thought that this behavior benefits prey capture, either through physically entangling the prey [53], or helping with sensing and identification of captured prey [2,26,27,50,52–58]. This behavior was sometimes called other names, including bouncing [32,48], thread pulling [51], shaking [47], jerking [50], plucking [52], and flexing [25] ), and web tensing [50,59]. To the best of our knowledge from reading these papers, they all correspond to dynamic crouching.

3. Shaking (Fig. 1(e), Movie S3 [165]). During prey capture [56], mating [47], intraspecific communication [51], or predator evasion [48–50], the spider performs rapid, repeated dynamic crouches with no pauses, resulting in high frequency dorsoventral body oscillations on the web, typically between 10–30 Hz [25,47–51]. At higher frequencies, the spider's body even pitches up and down substantially to result in fore-aft oscillations. It is thought that this motion helps establish dominance as it is used with the presence of other non-prey animals on the web [48,51]. Due to the similarity to dynamic crouching, this was sometimes categorized as the same movement compared to dynamic crouching. This behavior also was sometimes referred to as body jerking [47] and bouncing [48].)

Other leg behaviors such as plucking (picking and then releasing a thread with a front leg) and tapping (using a leg to tap a thread) occur, but only in non-prey contexts such as courtship or predator evasion [32,48] and thus are beyond the scope of this study.

As seen with many of the descriptions, every paper uses different terminology to refer to the same or similar spider motions. We have classified each motion to the best of our knowledge from reading these papers. To help put our work in the context of the literature, we created a figure to explain the important spider leg behaviors that we focus on and their relationship with the most relevant previous work (Appendix, Fig. 13).

We still do not fully understand the physical mechanisms of how spider leg behaviors contribute to web vibration sensing, largely due to the lack of quantification of vibrations with behaving animals. All the studies above on how web vibrations depend on web properties were done without a behaving spider on the web [2,22,27]. Similarly, most of the studies above that qualitatively described spider leg behaviors did not quantify their effects on web vibrations. In the rare cases where this was quantitatively studied, the spider was either replaced with an object of similar mass on the web [24,60], not actively moving on the web [27], or studied in isolation (while not on the web) [24,42]. Having a behaving spider (and a behaving prey, potential mate, or web invader) on the web would certainly affect the vibration dynamics. Thus, better understanding requires quantification of vibrations of the entire, coupled spider–web–target system with behaving animals.

This is hindered by major challenges faced by existing approaches. First, laser Doppler vibrometers commonly used in biological experiments can only measure at a single point at a time, because they use the reflection of a laser line via the Doppler effect to measure minute vibrations [2,27,42,61]. Thus, they cannot measure vibrations across the entire spider–web–target system with behaving animals, whose behaviors change the system state continuously. There are scanning laser doppler vibrometers that can scan multiple points but are very expensive and not practical for most biological research labs. These vibrometers also have difficulty recording high-amplitude vibrations as the laser line cannot track a largely oscillating web strand [27]. Recent techniques can use pixel intensity fluctuations high-speed videos to extract some features (but not the full 3-D motions) the entire web's vibrations [41,56,62], which alleviate this issue, but these studies did not quantify the full 3-D kinematics of spider behavior during prey sensing. Furthermore, the continual, inevitable changes in the web due to animal behavior and experimenter manipulation required to measure some key information can both preclude repeatable experiments. Web spiders regularly remake their webs and repair or modify the web (e.g., after damage by impacting or struggling prey, wind, rain, or morning dew) [14], and experimental measurements of web tension [63] and contamination causes the web to deteriorate quickly [14] resulting in disturbance or damage to the web. All these changes alter the web properties between trials and make it hard to tease apart effects from a single factor. Finally, although simulation studies using Finite Element Analysis (FEA) to model web vibrations [23,24,64,65] and using multibody dynamics simulations to model spider vibrations [42]) can address these challenges, adding behaving animals to the simulations multiplies their already high computational cost (from FEA applied to complex webs), and the lack of repeatable, quantitative biological data with behaving animals makes it difficult to validate such simulations.

To help alleviate these challenges faced by and add to insights from biological studies and theoretical/computation modeling, our long-term goal is to further establish the usefulness of the approach of robophysical modeling in understanding the physical mechanisms of how short-timescale spider leg behaviors contribute to vibration sensing of prey on an orb web. Robophysical modeling is the use of robots as active physical models to emulate and study biological systems in an experimental physics-like fashion [66–68]. Robophysical modeling has emerged as the third way (after theoretical and computational modeling) for modeling and understanding complex biological phenomena [68], thanks to several advantages. First and foremost, robophysical models operate in the real world and enact real physical interactions, whose complex physics may be missed by theoretical and computational models [66,69]. Furthermore, robophysical models can be made more amenable than organisms for systematic exploration of the parameter space with controlled, repeatable experiments and measurements of different types of information across system components [66,70]. Thus, robophysical modeling helps tease apart the effect of each parameter, allows easier testing of biological hypotheses and can even lead to new hypotheses [66,67,70–72], and facilitates a more wholistic understanding of how system dynamics emerges from interactions of its components [66,67,69,70]. A recent study already showed the usefulness of this approach in understanding how vibrations propagate through a vertical physical orb web [60], although how this was related to sensing was only investigated with a simplified passive rigid body spider physical model which did not have spider-like behavior [60]. Here we take the next step in using the robophysical modeling approach to help understand the role of spider behavior in vibration sensing of prey on a web, by creating a biologically relevant robophysical model that recapitulate the spider–web–prey system and using it to study how dynamic crouching may contribute to prey sensing using web vibrations.

Our creation and study of the robophysical model was guided by recent observations in a model organism, *Uloborus diversus* [73] (https://doi.org/10.1101/2025.06.08.658484) (Fig. 1). *U. diversus* is a nocturnal spider and belongs to the cribellate orb-weaver family Uloboridae [74–76]. It builds horizontal orb webs and hangs beneath the web, usually in the center in various positions. It dynamically crouches and shakes repeatedly as it moves towards prey caught on the web [43]. While the vertical orb is considered the "generic" form of a web, horizontal orb webs are widespread and ecologically relevant, spanning across multiple families of spiders (e.g. *Lecucauge*, *Uloborus*, *Metabus*) [77].

A recent study used spider leg movement to classify leg behaviors of *U. diversus* during prey sensing and capture [56]. A side view video of the spider was used to track spider anterior and posterior leg joints during prey capture. The frequency profile of each leg joint was used to train an unsupervised Hidden Markov Model, which categorized three behavioral motifs: static (~0 Hz), in which there is little leg movement; crouching (3–10 Hz), in which the spider slowly bends and extends its legs; and shaking (10+ Hz with prominent 10 Hz peak frequencies), in which the spider rapidly shakes the web [56]. These three behavior motifs do not correspond exactly with the behaviors as we define in this paper. A jerky crouch as we define here consists of a sequence of three behavior motifs starting from static, then to crouching, then to shaking, and finally to static. (Sometimes when jerky crouches are repeated, the static pose between each jerky crouch is extremely brief.) A soft crouch as we define here consists of a sequence of three behavior motifs starting from static, then to crouching, and finally to static.

As a first step towards our longer-term goal, in this study we focus on the role of dynamic crouching (specifically jerky crouch, as soft crouch does not induce substantial web vibrations; also see Section II.B.3). Among the three types of leg behaviors during prey capture (assuming different static postures, dynamic crouching, and shaking), dynamic crouching is most suspected to be a prey sensing behavior as explained above.

There are many hypotheses as to why spiders dynamically crouch on their web. The prevailing consensus is that a spider uses dynamic crouching to gain information about prey trapped on the web [2,26,27,50,52,53,55–57]. Spiders have been shown to use dynamic crouching to detect or locate various objects, ranging from vibrating prey [50] to non-moving objects like fly wings or weights [27]. It was speculated that, when the spider dynamically crouches, an object caught on the web can "bounce" a vibrational "echo" back to the spider, which informs the spider about its identity and location [26,50,57], although exactly how such a vibrational "echo" works remains unclear [26]. A recent study further hypothesized that the spider can use small-amplitude, high-frequency vibrations in the web to locate prey, similar to echolocation [2]. Specifically, the spider can pluck radial threads of the web to send a small-amplitude, high-frequency, transverse/lateral (perpendicular to the thread) wave of vibrations, which "bounce" back from an object as the object cannot oscillate so fast and causes a standing wave to form between itself and the spider [2]. This hypothesis leaves out the potential use of "echoes" related to low-frequency web vibrations, which can be present and involved during prey capture [20]. However, low-frequency web vibrations are generally less studied partly because the laser doppler vibrometer cannot reliably measure large vibration amplitudes accurately [27]. Regardless what frequencies may be involved, it remains unclear exactly how this vibrational "echo" mechanism works, nor is there clear physical evidence of this "echo" [2,26,53,55,78], due to the difficulty in obtaining repeatable, systematic, quantitative data of the spider–web–target system as discussed above.

Additional hypotheses include that dynamic crouching causes the prey to become more entangled in the web [53], that it induces prey to actively move which may generate vibrations to test whether an object in the web is alive or inanimate [54], and that it causes more tension in certain radial threads which allows the spider to focus its attention on certain web areas [58]. These hypotheses also lack a mechanistic explanation due to the difficulties in collecting physical data to demonstrate how this may benefit the spider during prey capture [2].

We aim to use robophysical modeling to discover physical mechanisms of how dynamic crouching *can* help our model organism detect prey using vibration sensing on a horizontal web, which may provide physical insights into and physical principles that govern some of the previous biological hypotheses or suggest new ones. We emphasize that our goal is **not** to uncover the neurophysiological mechanisms which will require neurophysiological measurements and is beyond the scope of this study.

To study the physical mechanisms that spiders *can* use when dynamic crouching to detect and locate prey, we created a robophysical model of the spider–web–prey system (Fig. 2(a)), consisting of a horizontal artificial web, a spider robot that can dynamically crouch under the web with vibration sensors on its legs, and a prey robot capable of actively vibrating (Section II.A). We then described how we analyzed the resulting leg and web vibrations (Section II.B and II.C). We first used this robophysical model to perform exploratory experiments to identify practical yet biologically relevant experimental treatments (Section III.A). This revealed that a single "jerky crouch" from the spider robot produced clear, interpretable signals that we design our main experiment around (Section III.B). In the main experiment, we systematically varied the prey robot's presence and location on the web (Section IV.A).

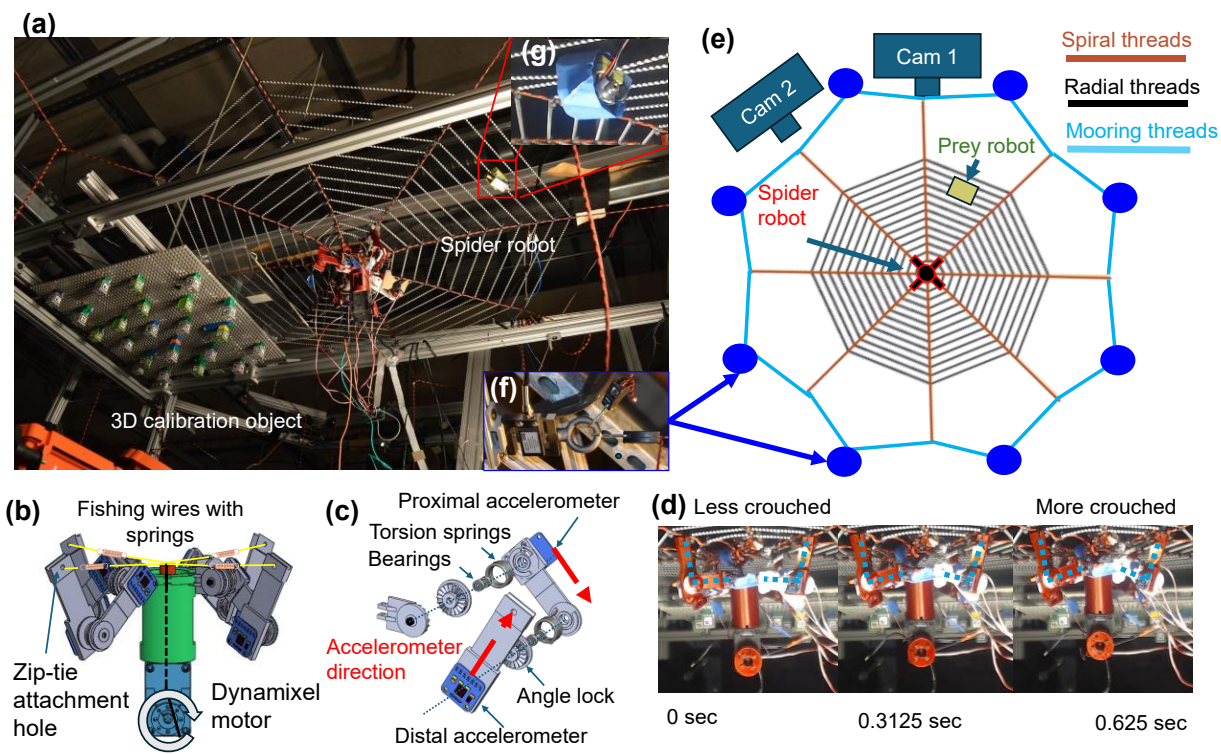

**Figure 2. Robophysical model. (A)** Photo of robophysical model. **(B)** Design of spider robot. **(C)** Spider robot leg design. **(D)** Schematic of robophysical model. **(E)** Example of load cells connected to mooring

threads. **(F)** A close-up of prey robot. **(G)** Snapshot of spider robot dynamically crouching.

Our main experiments revealed the physical mechanisms of how the spider may use dynamic crouching to detect prey on the web (Section IV.C). Without the prey robot, after a jerky crouch, the spider robot detected a dominant vibration frequency; with the prey robot, it detected an additional, higher dominant frequency (Section IV.C.1). In addition, this latter frequency decreased as the prey robot became farther from the center of the web at which the spider robot was (Section IV.C.2). Furthermore, our web vibration measurements showed that the former frequency was the natural frequency of the spider robot and the web passively vibrating, whereas the latter frequency was the prey robot's natural frequency on its spiral thread (Section IV.C.3). Because this spiral thread became longer as the prey robot became farther from the center of the web, this second natural frequency became lower. These findings meant that the spider *can* use a jerky crouch as a form of active vibration sensing to detect the presence and distance of prey on the web. We did not find conclusive evidence about how to detect prey robot direction (Section IV.C.4). Based on these robophysical modeling observations, we created a template model to explain the basic physical mechanisms (Section V). We also found evidence of the same physical phenomenon in videos of the *U. diversus* prey detection and capture on the web (Section VI). Finally, we summarize our work and discuss its implications, limitations, and future directions (Section VII).

## II. ROBOPHYSICAL MODELING METHODS

In this section, we first describe how we built a robophysical model to study the role of dynamic crouching (specifically jerky crouch) in prey sensing using cited literature (Section II.A). This is followed by general methodology to analyze the different vibrations in all our experiments (leg vibrations, Section II.B; web vibrations, Section II.C).

### A. Robophysical Model

The robophysical model (Fig. 2(a)) consisted of a spider robot, a physical web, and a prey robot.

The measurement and actuation devices were all connected to one data acquisition system to ensure that everything ran simultaneously.

**1. Spider robot design**

We created a spider robot (200 g weight) capable of dynamical crouching and sensing vibrations on a web. The spider robot design was simplified from the complex morphology of spiders [14], while still capturing the basic features of their vibration.

Spiders have two body segments and eight legs, each with eight segments and seven joints [14]. While each of the seven joints has a principal bending axis, many of them can also bend out of the plane defined by this principal axis [44]. The eight legs are arranged in lateral symmetry, and the four legs on each side differ in length.

For simplicity, our robot had a rigid body and four legs (all 3–D printed using PLA plastic), each with three segments and two joints (Fig. 2(b)). The choice of having only four legs was due to our design to use a single motor to actuate all legs; with this design, the motor was not strong enough to move eight legs. See discussion (Section VII.B) of further evidence from our recently created eight-legged, morphologically more accurate spider robot. The four legs were identical and arranged with radial symmetry. The two joints were near the tip and in the middle of each leg, respectively, resembling the coxa-trochanter and femur-tibia joints of spider legs. Each leg joint was confined to bending about a single axis, and the bending axes of the two leg joints were aligned, resulting in a main plane of bending, but passive give allowed slight bending of the whole leg slightly out of plane. To allow passive joint vibrations and movement of the legs, each joint had a torsion spring attached with hot glue onto a bearing inside the joint (Fig. 2(c)). Each leg tip was attached to the web through a pair of zip ties.

*a. Spider robot movement capabilities*

The robot dynamically crouched (Fig. 2(d), Video S4) by moving from a less crouched to a more crouched posture (joint angle changing from 20° to 40°) and then returning to the less crouched one. It could also lock itself into a less crouched or more crouched position. All the four legs were actuated to dynamically crouch in synchrony by pulling and releasing a fishing wire between the distal end of the leg and the dorsal end of the body via a servo motor (ROBOTIS Dynamixel XM430-W350-R) housed inside the robot body.

*b. Spider robot sensing capabilities*

To sense leg vibrations, we attached an accelerometer (ADXL326, (0.5–1600 Hz), range of acceleration of ±16 *g*, *g* being gravitational acceleration) near each leg joint, resulting in a total of eight accelerometers. While the lyriform organs typically resemble strain gauges [14], we found that accelerometers could also measure vibrations accurately. (This has also previously been demonstrated by another study [41,60].) We found that using accelerometers was a sufficient alternative to measure vibrations at the model web's scale. To measure the spider robot's leg vibrations, the eight accelerometers recorded accelerations along the direction of the distal leg segment (Fig.2(c)). To allow legs to still vibrate during leg actuation, a linear spring was added into each fishing wire. These accelerometers were connected to a 5V power supply for power and to a data acquisition system (USB-231, Measurement Computing Corporation) via thin, lightweight (26 gauge) wires to avoid affecting the vibration of the spider robot. Each accelerometer was labelled as proximal (P) or distal (D) with a leg number ranging from 1-4 (Fig. 4(a)).

**2. Physical web**

Our physical web design and construction followed those of a previous study which studied vibration propagation across a vertical web (but not the role of behavior in vibration sensing) [41,60]. Spider webs generally have two types of threads: high-stiffness radial (and mooring) threads for support, and low-stiffness spiral threads for prey capture [14]. We developed a horizontal physical web, with 8 radial threads attached to a ring of 16 mooring threads and 13 spiral threads between each adjacent pair of radial threads (Fig. 2(e)). We used stiff parachute cords (275 Paracord, Paracord Planet) for the radial and mooring threads (with black and orange stripes) and compliant shock cords (1/8" shock cord, Paracord

Planet) for the spiral threads (with black and white stripes). These cords have qualitatively similar viscoelastic properties to the radial and spiral threads of spider webs (specifically, the stress–strain relationships show a similar Young's modulus between the cords and their biological counterparts, but have different intercepts within the stress strain curve) [41,60]. The stripes of these cords could be tracked using computer vision to measure 3-D vibrations across the web threads. The spiral thread is strung around in an Archimedean spiral to a diameter of ~0.8m. The physical web weighed 720 g,

The web was attached on a frame made of 80/20 aluminum bars rigidly mounted to a ceiling and walls of the lab. The web was initially built in a loose conical structure on wooden frames, and then stretched outward, following methods in the previous work [60]. This was because as the spiral threads were put on the web, they gradually loosened the more spiral threads were put on a web, primarily on the outer edges due to the previous connections loosening over the course of construction. By stretching out the web from its conical structure to a flat structure, akin to opening the radial frame of an umbrella tightens its membrane surface, the web spiral threads all become equally stretched.

*a. Web tension monitoring*

We used tension monitoring to ensure that tension drift did not significantly affect our experiments. In the real spider's web, the web was pre-tensioned to control the vibration in the web [22]. Web tension can affect the vibration the spider receives [22]. This has also shown in a passive rigid-body spider physical on a vertical physical web [41]. To mitigate this, we added a load cell (Fig. 2(f), CALT DYLY103 30KG S Beam Load Cell Sensor, 2.0 ± 0.05 mV/V) to each of the eight anchors of mooring threads to monitor tension. We tightened each mooring thread (via a ratchet) to achieve about the same tension (3234 N ± 329 N) across all eight load cells. In preliminary experiments, we found that the web tension drifted over a long time (on the order of days on its own or after the spider or prey robot was removed and re-attached (even at the same location), resulting in changes in the vibration profile (Appendix, Fig. 14). Nevertheless, important features of the vibration dynamics central to our conclusions remain qualitatively unchanged (Appendix, Fig. 14). To mitigate this issue, all subsequent experiments were done in the span of 3 days.

### b. Web vibration measurement

We also developed a method to use high speed imaging to measure web vibrations in full three dimensions. First, we set up two high-speed cameras (Photron Mini UX100, 1280 × 1024 pixels, 1000 frame/s) to record a section of the web between two adjacent spiral threads, focusing on the stripes (spacing of every 3$^{rd}$ stripe) on each of the spiral threads. Then, we used the DLTdv8 digitizing tool [79] to track the stripes on the spiral threads in the videos of both camera views. In addition, we developed a semi-automatic computer vision tracking algorithm to identify the tracked stripes and match their identity between the videos of two camera views. Finally, we combined the tracked points from both camera views to perform 3-D reconstruction to obtain their 3-D coordinates using the direct linear transformation method [79]. To facilitate 3-D calibration, we built a calibration object using Lego bricks, with 88 BEEtag markers that can be automatically tracked [80], and mounted it to the web frame via a sliding system made from 80/20 parts to allow moving the calibration object in and out of the camera view repeatably between experiments.

### c. Web spiral thread numbering

In this paper, we varied the prey robot location by attaching it to different spiral threads. We denote each spiral thread by a number starting from spiral thread 1 being the closest spiral thread to the center of the web, and spiral thread 13 being the farthest spiral thread from the center of the web (Fig. 4(f)).

### 3. Prey robot

We created a prey robot (Fig. 2(g)) to emulate the dominant prey motion observed from our biological study in addition to others [27]. When the prey is captured, it behaves mostly as a solid rigid object, with occasional "struggling". The prey robot consisted of an appendage that moved relative to a body via a linear solenoid motor (12 V). The body was attached to a spiral thread of the web via to 3-D printed clamp. The prey robot weighed 90 g total. We attached weights (50 g) around the robot to make the prey robot heavier to make the vibration signals more salient. Each time the moving appendage moved, either from turning the solenoid "on" to "off" or vice versa, the prey robot actively vibrated (Movie XX).

4.  **Data acquisition system**

We developed a data acquisition system to record and synchronize all the data, including the actuation profile of the spider robot and prey robot that generate their active motions, passive vibrations of the spider robot legs from the accelerometers, passive vibration of the web from high-speed camera videos, and web tension from the load cells. A flow chart of the following description can be found in Fig. 3. To minimize the weight of those wires, we used fine wires (26 gauge) taped onto support structures with enough length to avoid affecting the robot's vibration.

To start the entire system, a trigger box (RW Electronics Standard Trigger Box) generated a rising-edge signal through BNC cables, which was sent to three devices. This included the two high-speed cameras that are set to record as soon as they receive the trigger signal, one DAQ (USB-231, Measurement Computing Corporation) that was responsible for measuring the spider robot's eight accelerometer signals, and another DAQ of the same model that was responsible for communicating between the trigger box and the control computer. This second DAQ prompted the computer to begin executing several armed MATLAB programs responsible for the two robot controls and web tension measurement. The prey robot's MATLAB code runs an Arduino UNO connected to a L298N to actuate the prey robot. The spider robot's MATLAB code runs a Dynamixel U2D2 connected to the spider robot's Dynamixel motor that runs a prescribed actuation profile (motor position vs. time) for the Dynamixel motor. Another code records tension measurements from the load cells from another Arduino UNO. The two synchronized high-speed cameras recorded web motion, and their data were sent back to the computer via Ethernet cables. The two high-speed cameras were run via PFV4. A web camera (Logitech C920 PRO HD Webcam) was used to record the experiment for documentation purposes.

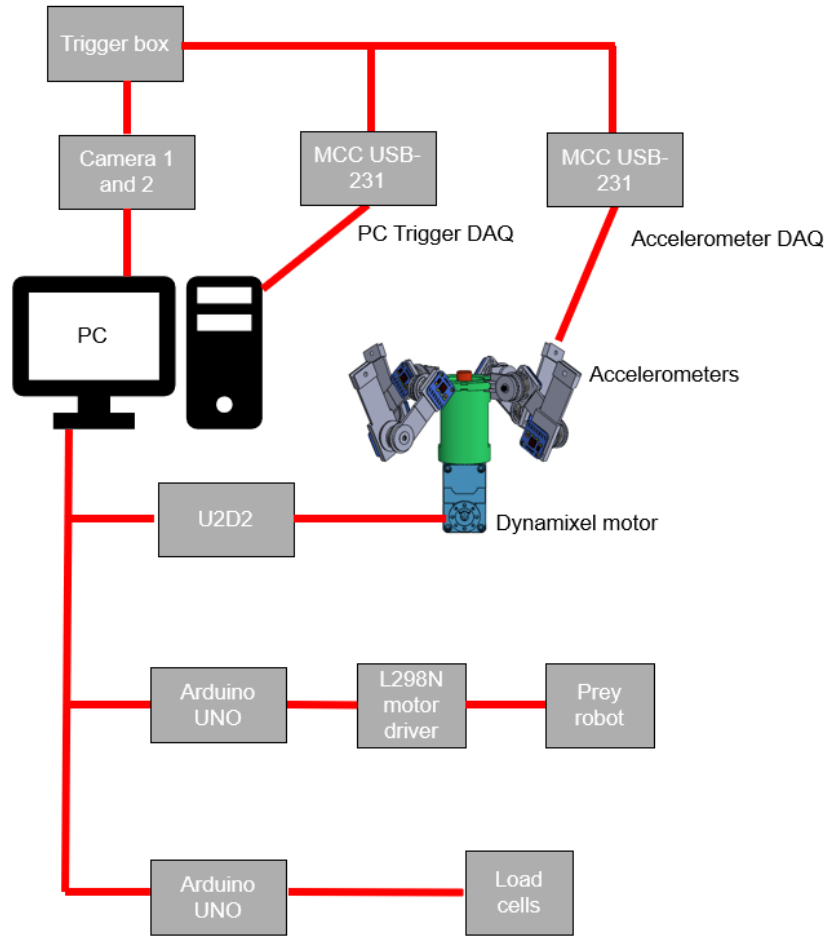

**Figure 3. A map of the robophysical model data acquisition system.**

B. **Vibration Analysis of Spider Robot Leg**

In our analyses of the vibrations in the spider robot legs, we used fast Fourier transform (FFT) to obtain the frequency spectrum of vibration data. The same general analysis methods were applied to all vibration-based signals, including: (1) accelerometer signals (voltage vs. time) measuring the spider robot's leg vibrations (this section), (2) 3-D positions vs. time of the tracked stripes of the physical web obtained from high-speed videos (Section II.C), and (3) 2-D positions vs. time of points of interest on the web from top camera video (Section VI.B) [56].

To analyze the vibration data for the robot experiments, we obtained the resulting vibration after the dynamic crouching movement of the spider robot, because when the spider dynamically crouched, the sensed leg vibrations were dominated by the spider robot movement, obscuring the resulting web vibration signals (as discovered in Section III.E.4). Any non-spider self-induced robot movement-related vibration was analyzed. The accelerometer data were recorded at 5000 Hz and filtered to a $6^{th}$ order Butterworth filter of 50 Hz to avoid the 60 Hz signal from electrical interference, and to smooth out noise within the accelerometers. There were no significant frequencies identified in all accelerometer data above 50 Hz before the Butterworth filter (other than three scenarios in the exploratory results, which we did not test in the main experiments; See Section III.B.5 and Section III.B.6).

In all treatments when the spider robot dynamically crouches, the data was cropped after the vibration data that starts after the spider robot finishes moving (Fig. 4(d)) This was defined as 0.02 seconds before the first large peak using the MATLAB find peaks function, after which 4.99 seconds (longer than a jerky crouch of the spider robot) was used for the Fourier transform data. This limit was chosen as the vibration signal had completely dissipated after 5 seconds. As a result, this signal aligned with the other signals from each trial (Fig. 4(e)). Then, the FFT was performed on the zero-mean data of each signal that was zero-padded (additional zeroes were added to the end of the signal to increase frequency resolution) to 6.55 seconds long, to increase the sample size so that the peaks were more obvious with the FFT algorithm. A 20 Hz 6th-order Butterworth filter was used instead for these treatments as there was no frequency identified above 15 Hz. In the treatments when the spider robot did not dynamically crouch, the 6 seconds of data was temporally cropped for analysis, starting from when the spider robot's accelerometers begin detecting vibration (Fig. 4(c)). How trials were averaged into a final plot for interpretation is described in Section III.B.1.

## C. Vibration analysis of web for dynamic crouching of spider

We analyzed web vibration data using a method similar to the accelerometer analysis (Fig. 3), extracting the 3-D vibration data out of points on the web through direct linear transform (DLT) from the footage of the two high-speed cameras recording the web (Fig. 6(a, b)). The web vibrations were only analyzed in the main experiments.

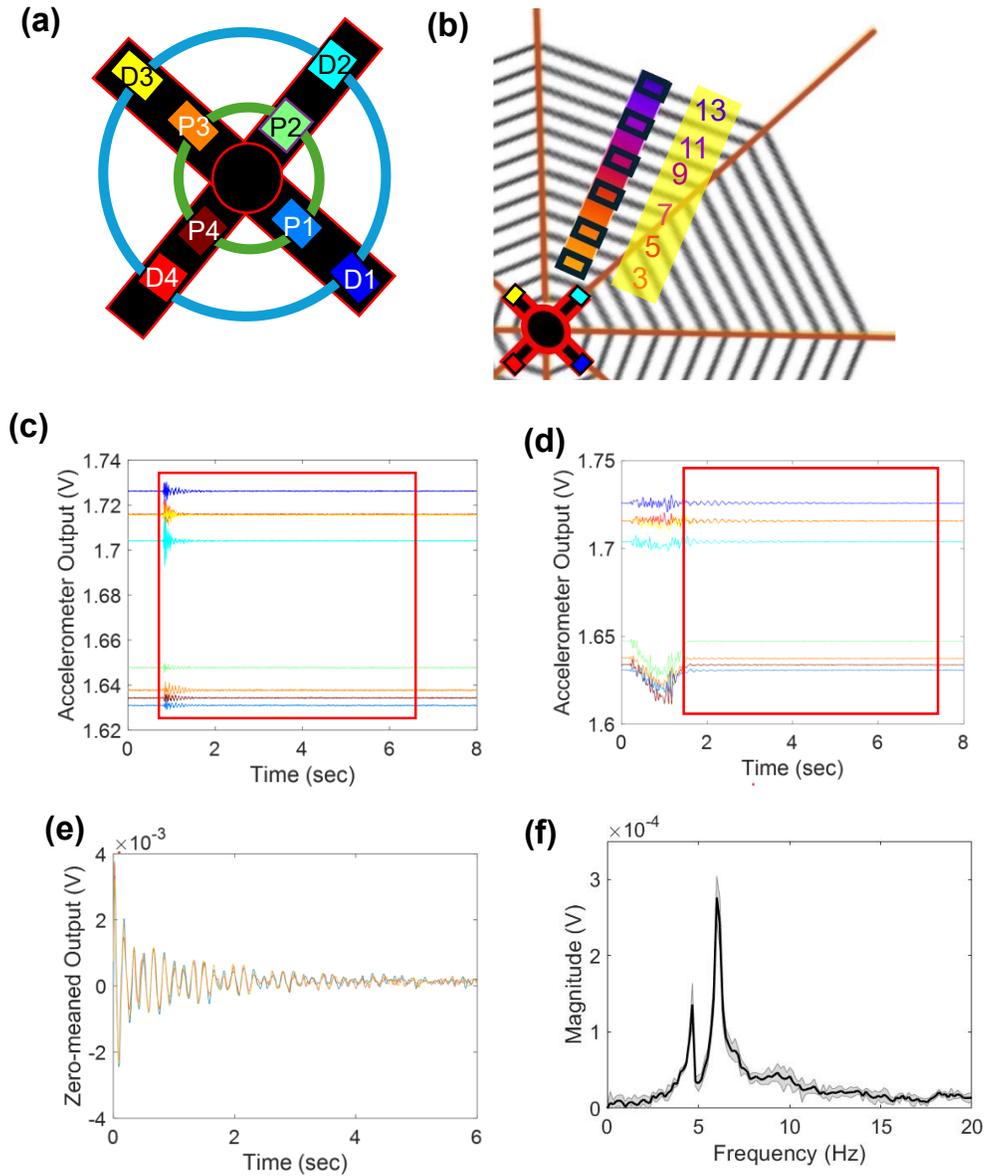

**Figure 4. Overview of the data analysis method. (a)** A map of the locations of each accelerometer. D represents distal, P represents proximal. **(b)** Prey robot location tested relative to the spider robot at the center of the web. The prey robot was in the sector of the web closest to Accelerometer D2. **(c, d)** Accelerometer output vs. time filtered through a 50 Hz 6$^{th}$ order Butterworth filter for the 8 accelerometers from one trial where (c) the spider robot was stationary and prey robot was actively shaking once and (d) the spider robot was moving and prey robot did not actively move. The red box indicates the data that was used for FFT analysis. The red box indicates the data that was used for FFT analysis, after the V shape indent, indicating where the spider robot had done its dynamic crouching motion. Each color represents the respective accelerometer identified in Fig. 4(a). **(e)** An overlap of a single accelerometer's raw data across 3 trials from Fig. 4(a). **(f)** The fast Fourier transform of accelerometer D2 with error bars (mean ± s.d.).

For all main experiments where the spider robot was dynamically crouching, we performed the same Fourier transform analysis as the accelerometers above for on the web vibration data of 3-D positions of the tracked stripes on spiral threads of one sector of the web obtained from the highspeed cameras. Due to camera field of view size limitations, we could only record the section of the web with the prey robot on it, and only spiral threads 1–9 of that section. We analyzed the web vibration data using a method similar to the accelerometer analysis (Fig. 3), extracting the 3-D vibration data out of points on the web through direct linear transform (DLT) from the footage of the two high-speed cameras recording the web (Fig. 6(a) and 6(b)). This was done through tracking the web vibration, the resulting videos are tracked using the DLTdv digitizing tool [79]. Both the spiral and radial threads are striped, and the stripes are then tracked using dltdv8a. We developed a semi-automatic computer vision tracking algorithm to detect these stripes and match them between the camera images. Finally, we reconstructed 3D kinematics of all tracked markers using the direct linear transformation method. To facilitate 3D calibration, we built a calibration object with 88 BEEtag markers [80] using Lego bricks (The Lego Group, Billund, Denmark).

Because the spider robot was moving up and down on the web, shaking the web transversely, we only extract the transverse direction (the axis parallel aligned with the direction of gravity) to analyze the vibration. We only extract the portion of data when the spider robot was not moving, and perform a fast Fourier transform for each point on the web. We then examined the Fourier transform of the transverse direction and, at each point, calculated the magnitudes of the different frequencies observed on each spiral thread, averaging them (Fig. 8(c) and 8(d)). We also analyzed the longitudinal (along the thread) and lateral vibrations (within the web plane, perpendicular to the thread), but we found that those directions were noisy and difficult to interpret, as the spider robot did not vibrate substantially in these directions.

## III. EXPLORATORY EXPERIMENTS

Because this was the first time robophysical modeling has been applied to vibration sensing, we performed exploratory experiments to identify practical yet biologically relevant experimental treatments to test in our robophysical model and to establish data analysis protocols for the resulting data (Section III.A). These tests revealed key insights of our robophysical model (Section III.B) that we then used to design and perform a main set of experiments where we varied prey robot presence and location (Section IV.A). We also hope that our documentation of exploratory experiments (Section II) provides a good example of how such robophysical modeling was and should be done in early stages of a study (especially on a topic little explored using this approach, such as vibration sensing), as this approach is less widely used than theoretical and computational modeling.

### A. Exploratory experiments to seek a biologically relevant focus experiment

Because this was the first study to use robophysical modeling to understand the role of leg behavior in vibration sensing in spiders, there was no established experimental design and protocols to follow. In the biological system, the spider does jerky crouching, soft crouching, shaking, and intermediate motions between them, combined with locomotion in quick succession [56]. This complicates the isolation of the behavior that we would like to observe—jerky crouch—so that we can model the same behavior in our robophysical system. In addition, it was initially unclear how to analyze our robophysical model's data. To address these, we first performed a series of exploratory experiments to identify a specific "jerky crouch" motion profile that produced a consistent, interpretable vibration signals, and used the results of these exploratory experiments to inform the design of our main experiments.

The methods of the exploratory experiments were not systematically repeated for all the treatments explored, unlike as was done in the main experiments, but they were nonetheless crucial for informing how we designed the main experiments. In exploratory experiments, we tested: the types of motion profiles for the spider and prey robot, and viable combinations of these motions that could be realistically and repeatably modeled in the robophysical system. These allowed us to select the spider robot's and prey robot's motions appropriate for systematic study using our robophysical model and for later comparison with biological observations. In our exploratory experiments, we only analyzed the spider robot's accelerometer data to see how each treatment affects the leg vibrations of the spider robot and how these vibrations should be analyzed, but we did not record the web vibrations (which we later added in the main experiments in Section IV.A). Some motions that were chosen to be tested, did not represent the jerky crouch behavior we wanted to test (e.g., a spider robot not actively moving), or did not represent biological observations (e.g., a spider does not actively crouch when there is no prey on the web) but our learnings help us justify how we chose the main experiments.

**1. Spider robot treatments to identify how spider robot should move**

In the parallel biological study (https://doi.org/10.1101/2025.06.08.658484) [56], consistent with the broader biological literature, the spider does an assortment of behaviors during prey capture (Fig. 1). We tested what spider robot movements best represented dynamic crouching, specifically jerky crouch, as later we find that soft crouching did not produce any vibrations within the web. We wanted to find a spider robot movement that balanced biological realism and robophysical modeling feasibility.

*a. Repeated crouching of spider robot*

The spider can do a repeated jerky crouch behavior during prey capture [56]. To emulate this behavior, we tested the spider robot repeatedly crouching. This motion involved repeatedly going from a less crouched, to more crouched, back to a less crouched posture with a cycle period of 1.25 seconds. This cycle frequency of 0.8 Hz was the maximum that the spider robot could achieve.

*b. Single crouch of spider robot*

Another behavior of interest was a single jerky crouch (Movie S1 [165]). In this robot treatment, the spider robot crouched only once and remained static afterwards. The motion only goes from less crouched to more crouched to less crouched in the span of 1.25 seconds (Movie S4 [165]). This resembles the jerky crouches that we see in biology, as after the spider robot did this motion, the web was induced to vibrate.

c. *Static spider robot in less vs. more crouched posture*

Previous work hypothesized that the spider assuming different static postures functions as a reconfigurable mechanical filter to help the spider tune in to specific frequencies [42,46]. We tested whether the static spider robot can detect a moving prey robot and if positional adjustments were useful for detecting a moving prey robot. In this treatment, the spider robot was static and in either the less crouched or more crouched position.

**2. Prey robot treatments to identify how prey robot should move**

Prey captured in a web typically do not buzz (which produces high frequencies) but rather produce highly irregular vibrations below 50 Hz [26,33]. We tested how the minimalistic prey robot could shake to better represent this. In all treatments of this section, the prey robot was hung on the 7$^{th}$ spiral thread.

a. *Prey robot actively shaking continuously*

Here, we tested whether continuous oscillations of the prey robot can produce highly irregular vibrations under 50 Hz that are representative of a prey robot. Here, the prey robot moved at 12.5 Hz by turning the linear solenoid on and off rapidly during the entire trial. This movement caused the web and the prey robot to shake strongly.

b. *Prey robot actively shaking once*

Based on our observations and other literature [50], sometimes the prey does not move much, and the spider typically does the jerking motion in response to an unresponsive prey. Here, we tested where the prey robot moved only produced one light impact, by actuating the prey robot solenoid from "on" to "off" once. This is referred to as a "weak" motion, due to the light impact (Movie S5 [165]).

*c. Prey robot impacting web*

A previous study found that the spider can quickly orient themselves (<1 sec) and find the direction of the prey after it impacts the web without needing to dynamically crouch [26]. Here, we used a thread to hang an additional weight of 500 g to the prey robot (50 g) attached to the web. After the weight became stationary, the string was cut. The prey robot immediately bounced upward due to the elasticity of the spiral thread (shock cord), and a large amount of vibration was produced, inducing a prey robot motion, emulating the large amplitude web vibrations after a prey impact that gradually damps out. This is referred to as a "strong" motion, due to the heavy impact of the prey robot onto the web.

*d. Prey robot not moving actively*

Some studied found the spider can detect and locate objects not actively moving, such as detritus, immobile prey, or weights [26,28,81]. Here, we simply hang the prey robot on the web but did not actuate it to move actively. It could still move passively, induced via the web motion by the spider robot.

*e. Prey robot not on web*

We also tested the case with the prey robot not on the web, as a control treatment.

**3. Testing repeatability of robophysical model**

Not all combinations of prey robot and spider robot motion were tested due to certain motions of the prey robot and spider robot determined as unviable before analysis. In those particular cases, due to the difficulty of analysis, we only perform analysis on one trial of these treatments. This was usually the product of one or both robots vibrating excessively and chaotically and thus it became hard to align these vibrations with the signals recorded from other trials. Once we identified the viable treatments to test for the systematic main experiments, we performed three trials for these exploratory experiments and assessed how repeatable our robophysical modeling was. This also allowed us to determine how many trials we needed in the main systematic experiments.

## B. Insights from exploratory experiments

We explored different variations of what the robophysical model could model by changing how the spider robot and prey robot behaved. Not every combination of spider robot treatment and prey robot treatment was explored, as either there was nothing significant to record (nothing is moving on the web) or a combination was already ruled out from the results of another combination. A summary of the results of the Fourier transforms of the different variations analyzed are in Fig. 5.

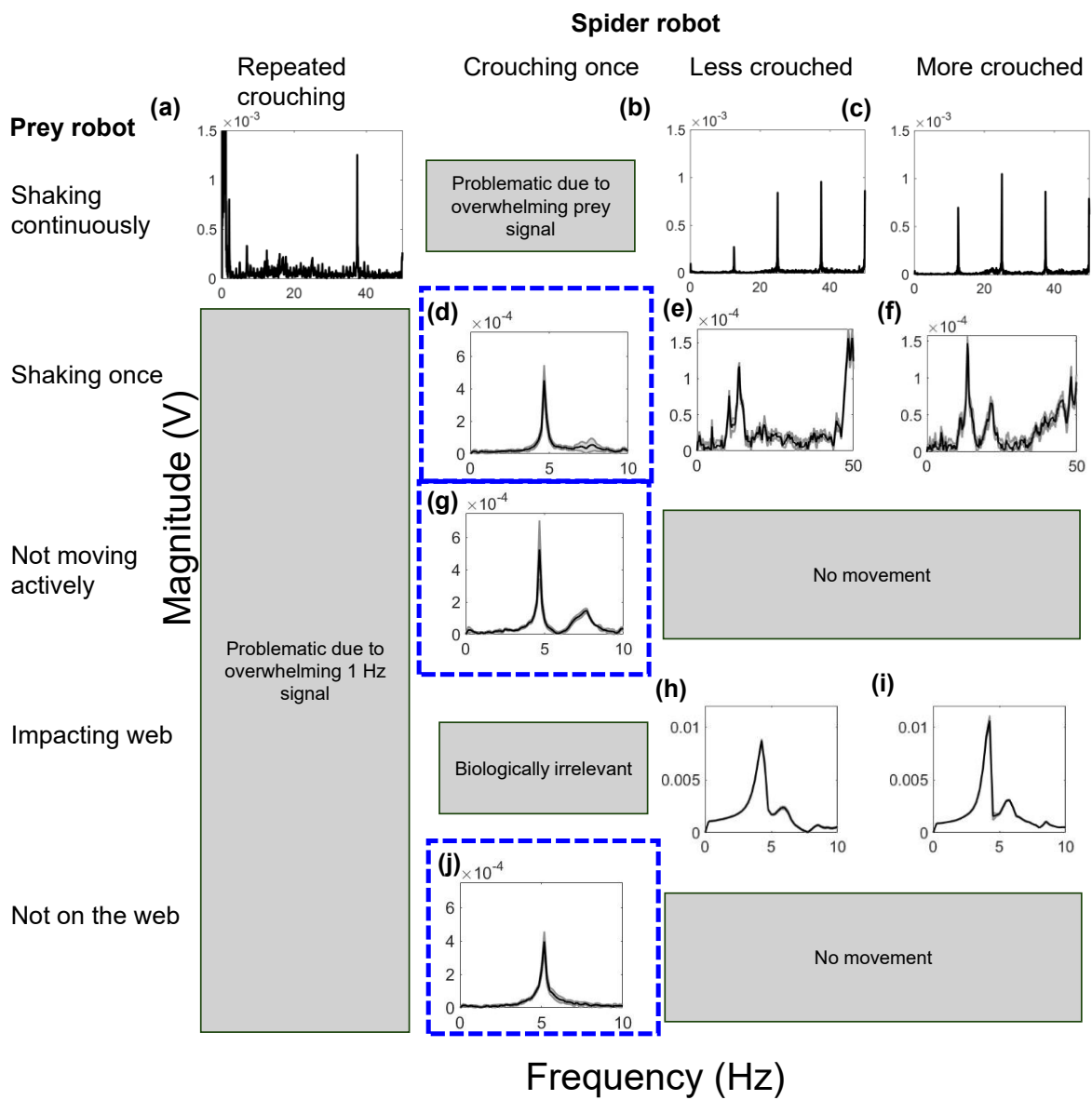

**Figure 5.** A summary of Fast Fourier Transforms (FFTs) from the exploratory experiments to help narrow down design of main systematic experiments. Each FFT compares magnitude (V) vs frequency (Hz). Except for A, B, and C, the FFT was the average between 3 trials. The blue highlighted examples are further studied in the main experiment. **(a-c)** FFT of treatments tested when prey robot was continuously moving. **(d-f)** FFT of treatments where prey robot moving once. **(g)** FFT of spider robot crouching once when prey robot was not moving. **(h-i)** FFT of treatments where prey robot impacts the web hard. **(j)** FFT of spider robot crouching once when prey robot was not on web.

1. **High repeatability of robophysical model**

    In most treatments, other than the prey robot continuously moving (Fig. 6(a)) and the spider robot dynamically crouching continuously (Fig. 6(b) and 6(c)), the resulting spider robot vibrations from multiple trials (after temporally aligned to start from the highest peak of the vibration, indicating the start of the post-spider movement vibration signal) basically overlapped (Fig. 4(e)), with minimal variation. An example of how well the acquired signals from each trial overlap each other can be shown in Fig.4(e), showing excellent repeatability of the robophysical model. We performed the fast Fourier transform (FFT) aggregating from the three trials per treatment to understand the frequencies within the spider robot's eight accelerometers (Fig. 5(e)). The vibrations from the three trials are almost identical if the experiments are done in quick succession in Fig. 5(c). As a result, in the systematic main experiments, we only collected 3 trials per treatment. To average the resulting Fourier transforms, the mean and standard deviation of the absolute value of the FFT data in the complex number form were plotted to obtain the averaged FFT plots (Fig. 4(e)).

2. **Distal accelerometers show similar vibration profile**

    In all these exploratory experiments, we found that all distal accelerometers closely resembled each other. Thus, only one accelerometer was analyzed (accelerometer D2) closest to the prey robot (Fig. 4(a)). This will also be the closest accelerometer to the sector of the web we analyze in Section IV.A. The

proximal accelerometers had weaker vibrations that were qualitatively similar and were not analyzed within the paper. The vibrations in the proximal accelerometers were weak such that some frequency peaks that existed in the distal accelerometers were too small to identify.

### 3. Soft crouch does not produce any vibrations

During exploratory experiments, we tried to generate the soft crouch, by having the spider robot dynamically crouch slowly up and down once. However, when our robot dynamically crouched once, it ended up vibrating the web significantly due to the sudden stop when it finished moving. This resembled the jerky crouch observed in the real spider (Fig. 1(c), Movie S4 [165]). For our spider robot to generate a soft crouch, it would have to move at a significantly slower speed, which would not vibrate the web substantially afterwards. Therefore, we only further investigated dynamic crouching with a fast actuation profile (i.e., jerky crouch) in subsequent experiments.

### 4. Repeated dynamic crouching overwhelms prey information

When the spider robot repeatedly crouched, the spider robot's movement dominated the leg vibration signals when it was jerky crouching, with a strong 1 Hz signal corresponding to its crouching frequency (seen by the 1 Hz peak in (Fig. 5(a))). This overwhelms and masks the vibrations from the prey robot, making it difficult to extract information about prey vibration data. In other treatments where the spider robot dynamically crouched (e.g., spider robot does a jerky crouch with a stationary prey robot), its motion would also result in a dominant 1 Hz frequency peak in the leg vibrations sensed by the spider robot, and thus there were no other subsequent peaks to identify in the Fourier transform.

However, once the spider robot stopped actively moving, the subsequent leg vibrations it experienced from the vibrations of the web clearly showed the prey data influence by showing clear differences in the peaks observed in the FFTs (described more in Section III.B.7). This result indicated that a pause after the spider's dynamic crouching is essential for the detection of prey. This was supported by the observations of the parallel biological study that the spider never repeatedly shakes and immediately locates and reaches the prey location [56]. As a result, we did not further investigate other treatments with repeated spider crouching. In addition, all subsequent analysis of the vibration data were done when the spider robot did not actively crouch. For all treatments where the spider robot actively crouched once (Column 2 in Fig. 5), we focused on analyzing only the resulting vibration of the spider robot after a jerky crouch (e.g., Fig.3(d)). When the spider robot did not actively move at all (stayed in a less or more crouched posture), the entirety of the vibration is analyzed. More detail on how this was done can be found in Section. II.B.

5. **Prey robot should not actively shake continuously**

When we tested the prey robot actively shaking continuously, the prey robot vibration was found to be overwhelming and caused subsequent passive vibrations to be dominated by the frequencies and the harmonics that the prey robot was oscillating at. When the prey robot was actively and continuously shaking at 12.5 Hz, there were strong peaks at 12.5 Hz, 25 Hz, 37.5 Hz, and 50 Hz in the fast Fourier transforms (Fig. 5(b) and 5(c)). Subsequent harmonics were less visible due to the 50 Hz Butterworth filter that obscured all vibrations afterwards. On the other hand, when the prey robot only actively shook once, the leg vibrations of the static spider robot showed highly complex signals, with many small peaks of 5 Hz, 12 Hz, 15 Hz, 22 Hz, and a broad range of 30+ Hz (Fig. 5(e) and Fig. 5(f)). The vibrations were removed after 50 Hz due to the Butterworth filter to remove noise, as they did not exhibit significant frequencies before the filter was applied. While the active continuous prey robot movement did not represent the random irregular 50 Hz vibrations observed by many biologists on the prey [26,33], the single active shaking of the prey robot more closely resembled the random sub-50 Hz vibrations, and visually represents the movement observed by prey on the web, where it only occasionally and randomly struggles on the web. As a result, further research was not conducted when the prey robot was actively shaking continuously.

Vibrating the prey robot at prescribed frequencies resulted in overpowering vibrations with those frequencies only expressed in the web, which was uncharacteristic of prey capture. As a result, we did not further test when the prey robot was continuously shaking but rather focused on when the prey robot only shook once. While live prey often struggles intermittently, modeling this as a single, discrete perturbation allowed us to generate a consistent prey robot vibration response that did not present the same issues as the continuous vibration response. This impulse response approach resembles previous biology experiments where a square wave was inputted into the web as a vibration stimulus [22,23].

## 6. Static posture does not affect vibrations significantly

Whether the spider robot was less crouched or more crouched, the sensed frequencies were similar, with only a major difference in magnitude (Fig. 5). This was visible in the static posture treatments where: (1) the prey robot shook continuously (Fig. 5(b) and Fig. 5(c)), (2) the prey robot shook once (Fig. 5(e) and Fig. 5(f)), and (3) the prey robot has impacted the web (Fig. 5(h) and Fig. 5(i)).

While the spider robot was statically crouching and the prey robot was continuously shaking (Fig. 5(b) and 5(c)), both postures resulted in the spider robot's legs vibrating at the 12.5 Hz and its harmonics that the prey was shaking at. However, the magnitudes of each frequency differed between the more and less crouched spider robot posture treatments. The spider robot sensed an overall larger magnitude of vibration when it is more crouched (Fig. 5(c)). However, due to the strong, identifiable peak frequencies and their harmonics, which did not resemble biological prey vibration, this was not further investigated.

After the prey robot shook once while the spider robot was not actively moving, the prey robot produced a multitude of irregular vibrations under 50 Hz, somewhat similar to the random sub-50 Hz vibrations observed in biological studies [26,33]. However, the difference of static leg postures did not modify the peak frequencies nor their magnitudes significantly (Fig. 5(e) and 5(f)). (Additional plots where the prey robot was put in different spiral threads can be found in Appendix, Fig. 15).

After the prey robot impacted the web strongly while the spider robot was not actively moving, the spider robot sensed two peak vibrations of distinct frequencies (Fig. 5(h) and 5(i)). However, the static leg posture did not affect the peak frequencies nor their magnitudes significantly. A previous simulation study inspired by the black widow spider's different postures during different states (e.g., more crouched, neutral) found that a spider's static posture affects the frequencies and magnitudes of the vibrations sensed by spider leg joints for a given vibration input at the leg tip, i.e., by assuming different postures, the spider's body and legs act as a reconfigurable mechanical filter [42]. However, in all our treatments, we were not able to observe this phenomenon, which may be due to the relatively small range of motion of the spider robot in changing its leg posture. Thus, we did not conduct further tests investigating the effects of static crouching.

### 7. A jerky crouch creates a low-noise vibrations with one or two peaks

After the spider robot performed a jerky crouch, it sensed a low-noise, transient response where distinct frequency components were visible (Fig. 5(d) and 5(g)). This was unlike the other vibrations that involved the spider robot continuously shaking or not actively moving, where there were many hard-to-identify frequency components. Enabled by the clean signal sensed by the spider robot, we find that when the spider robot jerky crouched with a prey robot either shaking once (Fig. 5(d)) or not actively shaking on the web (Fig. 5(g)), there were two frequency peaks in the subsequent vibrations sensed by the legs. (While the second frequency peak appeared less distinct in this trial (Fig. 5(d)), the two peak-structure was clear and robust when the prey robot was moved to other spiral threads in our systematic main experiments; see Section IV.C.1 and Fig. 6). On the other hand, when the spider robot performed a jerky crouch with no prey robot on the web (Fig. 5(j)) there was one clean frequency peak.

Compared to the treatment where the prey robot was actively shaking once and the spider robot was not actively crouching (Fig. 5(e) and 5(f)), while the prey robot could be detected through irregular sub-50 Hz vibrations, the frequency peaks were much less clear. This suggested that the spider dynamically crouches to get a clearer signal from the prey robot to locate and identify it, which was observed in the biological study [56]. It may also prompt the spider to do dynamic crouching motions to locate both moving and non-moving prey, as seen in biological observations [27,50]. We speculate that, if the prey robot moves once, the spider robot can use dynamic crouching to induce stronger vibration signals that can reveal more information about the prey robot than if it the spider robot was static or dynamically crouched continuously. Thus, we focused on this combination of spider robot and prey robot treatments in our main experiments.

## 8. Large impact of prey robot creates similar vibrations to when spider robot was dynamically crouching

Similarly, when the prey robot impacted the web (Fig. 5(h) and 5(i)), the leg vibrations sensed by the spider robot not actively moving were very similar to when the spider robot dynamically crouched once, with two vibration peaks (Fig. 5(g)). This suggested that, if a prey shakes the web intensely at impact, the spider may receive clear information about the prey without dynamically crouching, matching the observations that the spider immediately responds to prey (and only used dynamic crouching to detect the signal again) in the parallel biological study [56]. This was tested as an additional treatment that will not be focused on in our main experiments as we are primarily interested in the effects of spider dynamic crouching. (Additional results where the prey robot switched locations on the spiral thread can be found in Appendix, Fig. 16).

## IV. ROBOPHYSICAL MODEL MAIN EXPERIMENTS

### A. Robophysical Model Main Experiment Design

Informed by our results from our exploratory experiments, we designed our main experiments. We focused on using the robophysical model to study the spider's jerky crouch on the web and tested whether the spider robot can use a jerky crouch to detect the prey robot's presence, distance, and direction.

We conducted experiments to test whether a single jerky crouch (Section III.A.1.b) can help the spider robot detect the presence (Section IV.C.1), distance (Section IV.C.2), and direction (Section IV.C.4) of the prey robot. In these main experiments, we had the spider robot perform a jerky crouch on the web with: no prey robot (Section III.A.2.e), a not actively moving prey robot (Section III.A.2.d), and a prey robot that actively shakes once (Section III.A.2.b).

When the prey robot was on the web, we also varied the distance of the prey robot, by hanging it to different spiral threads (3, 5, 7, 9, 11, and 13) in the sector of the web where the cameras were pointed to (Fig. 2(e)). We collected 3 trials for each treatment (1 no prey treatment, 6 prey robot not actively moving treatments, 6 prey robot actively shaking once) resulting in a total of 39 trials.

B.  Robophysical Model Main Experiment and Analyses Protocol

The main robophysical experiments and analyses protocols are the same as those of the exploratory experiments (see Section II.B). In this case, the web vibrations were also analyzed (See Section II.C).

In the treatment where the prey robot was moving, the prey robot was timed to move 0.25 seconds after the spider robot began dynamically crouching. The prey robot's solenoid moved from the "on" position to the "off" position to make the single weak impact. When the prey robot was shifted between different spiral threads, we moved the prey robot to the center of the spiral thread for each treatment.

Additionally in these treatments, web vibrations are also investigated along with the accelerometer vibrations and are compared to ensure that the accelerometers are properly representing the vibrations appearing in the web. How web vibrations were analyzed can be found below in Section II.C.

C.  Robophysical Model Main Experiment Results

This section includes four sections that show four major findings: (1) a jerky crouch can help detect prey presence very clearly by a distinct, added vibration frequency; (2) the distinct frequency produced after a jerky crouch encoded the prey robot distance;(3) the frequencies in the web after a jerky crouch relate to the overall system and prey robot's natural frequencies; and (4) it was difficult to identify direction of the prey robot using our much simplified spider robot.

1.  **Prey robot presence can be detected by an additional frequency**

As mentioned in the earlier exploration results in Section III.B.7, we discovered distinct features in the leg vibrations of the spider robot after a dynamic crouch that encoded whether there was a prey robot on the web (Fig. 6). This involved two frequencies in the vibration where there was a prey robot on the web, and only one frequency in the vibration when there was no prey robot on the web. This was found to be mostly (10 out of 12 treatments) true when the prey robot was in other spiral threads. These observations also held for the distal leg joints of other legs that were farther away from the prey robot (consistent with our findings in Section II.E.1), although proximal leg joints did not show such distinct trends (Appendix, Fig. 17 and Fig. 18). As a result, we only focused on accelerometer on the distal segment of the leg closest to the prey robot (D2) throughout the rest of the results, as discovered in Section II.B.2.

Without a prey robot on the web, after a dynamic crouch, there was a distinct high peak at 5.3 Hz in the frequency domain (Fig. 6(a)). With a not actively moving prey robot on the web, a second, distinct, higher frequency appeared in the leg vibrations (Fig. 6(b)). If the prey robot was too close, such as in spiral thread 3, only one frequency was observed, though it differed from the no prey robot treatment's frequency at 5.3 Hz (Fig. 6(b)). There may be a second frequency observed in this condition, but it was too small to observe and can be mistaken for noise. (In Section V.B, we will confirm that this was just a very small second frequency.) However, the first peak frequency was at 4.6 Hz, lower than the 5.3 Hz that the spider robot detected when there was no prey robot on the web.

A prey robot shaking once did not change the frequencies of two aforementioned peaks and only reduced their magnitudes (Fig. 6(c)). The irregular sub 50 Hz vibrations from the prey robot shaking once and spider robot static treatment (see Fig. 5(e) and 5(f)) did not appear anywhere in the vibrations; they were completely overridden. Thus, the spider robot could still detect it by assessing whether a second frequency was present.

Therefore, after a single dynamic crouch, by assessing whether a distinct higher frequency was present in the leg vibrations than the distinct, lower frequency, the spider robot could detect whether there was a prey robot on the web. This applied whether the prey robot was shaking once or when the prey robot was not shaking, potentially applying to both moving prey and small debris for the real spider.

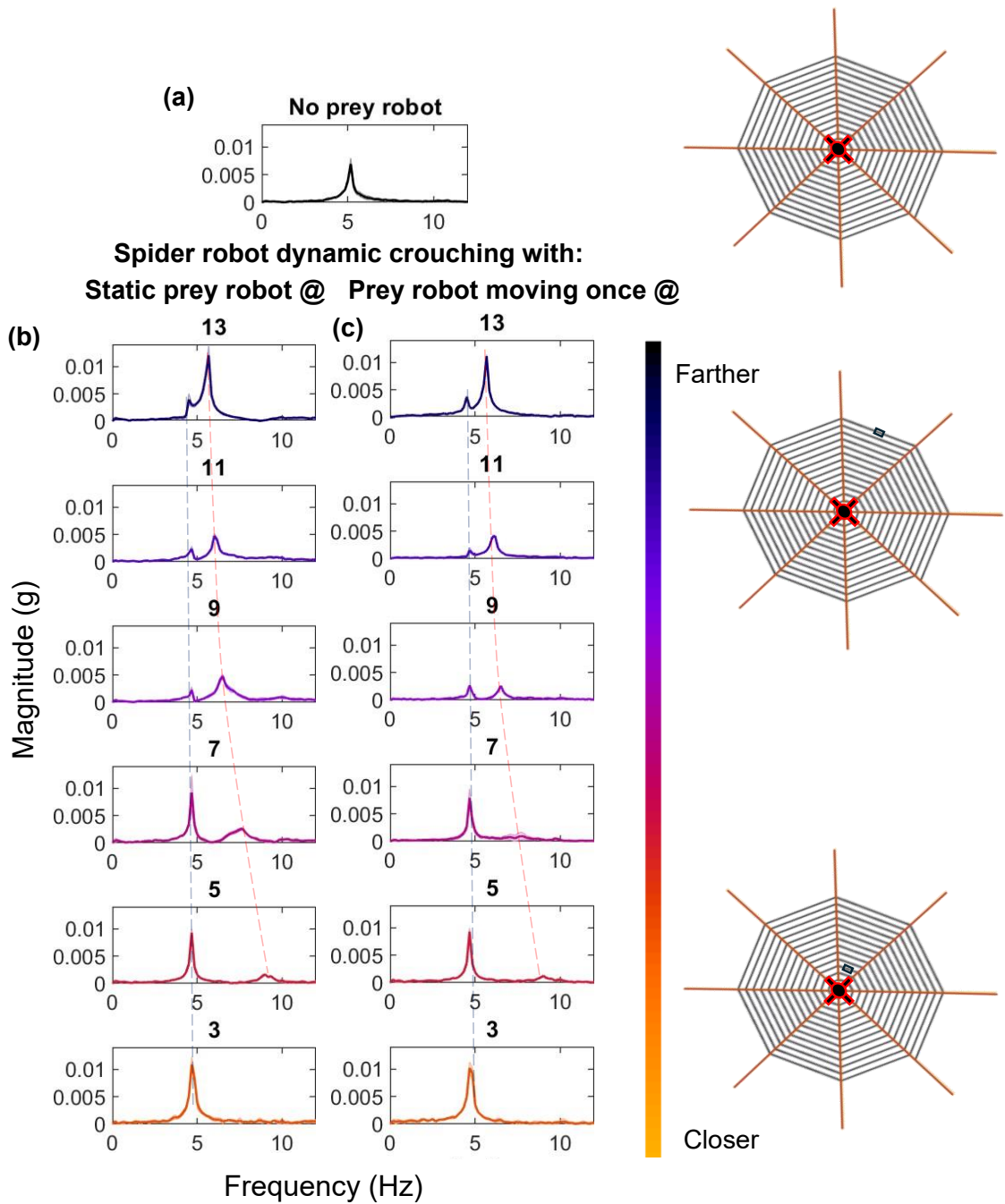

**Figure 6. Vibrations of spider robot legs after a dynamic crouch.** Data shown for accelerometer D2. See Appendix, Fig. 17 and Fig. 18 for data of all legs. The FFT's are averaged across 3 trials, with error bars (mean ± s.d.). Due to the extremely small error values, they are not fully visible (a closeup is shown in Fig. 3). **(a)** Frequency vs. magnitude (volts) without prey robot, from Fourier transform of acceleration

(volts) vs. time. **(b)** Frequency vs. magnitude with static prey robot. **(c)** Frequency vs. magnitude with prey robot moving once. In (b) and (c), top to bottom rows show data with prey robot on spiral thread 13 through 3. Blue dashed line shows natural frequency of the system. Red dashed curve shows prey robot's natural frequency.

**2. Prey robot distance is encoded in the additional frequency**

We also found that this additional higher frequency was correlated with the prey robot's distance to the center of the web. As the prey robot moved closer to the spider robot, from being on spiral thread 13 to being on spiral thread 3, the frequency of the added vibration peak monotonically and substantially increased from 5.6 to 10.2 Hz (Fig. 6(b), 6(c)), while the magnitude of the peak decreased, both for a stationary prey robot (Fig. 6(b)) and for a prey robot that moved once (Fig. 6(c)). These observations also held true for the distal leg joints of other legs that were farther away from the prey robot, although proximal leg joints did not show such distinct trends as their two peak frequencies were less distinct (Appendix, Fig. 17). This higher frequency was likely the natural frequency of the prey robot passively vibrating on the spiral thread that it was attached to, induced by the shaking of the web after a dynamic crouch and thus will be referred to as the prey frequency.

In contrast, the frequency of the lower, original vibration peak (4.5–4.7 Hz) did not change significantly in cases when the prey robot was on the web (Fig. 6(b), 6(c)), while its amplitude increased, as the prey robot moved closer to the center of the web. This lower frequency of around 4.5 Hz was likely the natural frequency of the spider robot and the prey robot coupled to the web passively vibrating in unison and thus will be referred to as the system frequency. This frequency was defined by the total effective mass (with slight variation due to the location of the prey robot) of the spider-prey-web system assembly. When the prey robot was not on the web, this frequency increased (likely due to the entire web becoming lighter) to 5.3 Hz. Because of this, we can conclude that (1) the system frequency was the baseline resonant frequency of the entire system, and (2) this provides an additional indicator of prey presence through its frequency shift. As described in Section IV.C.1, even if the higher frequency was too small to be detected, the spider robot still can identify prey robot presence, by comparing the system frequency to the no-prey baseline.

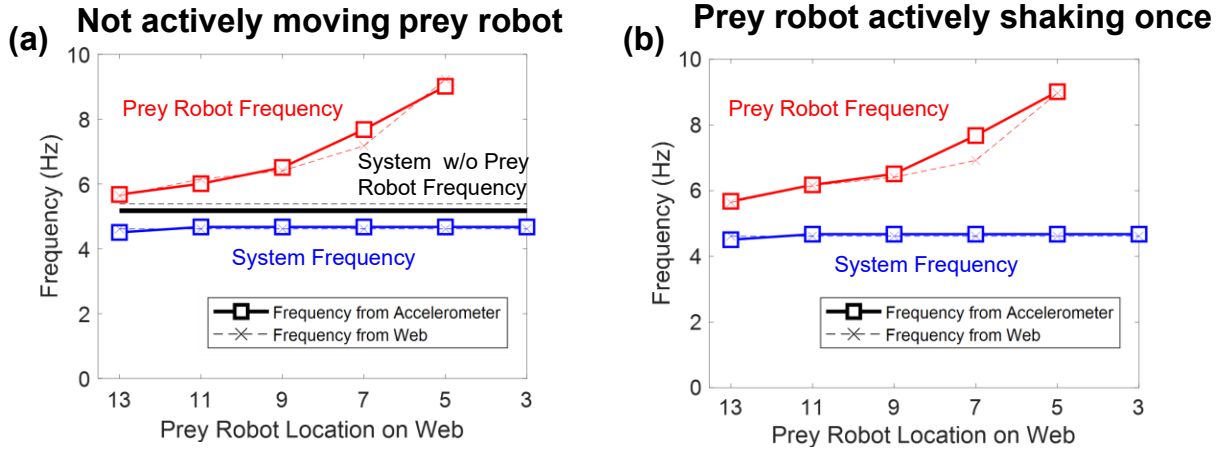

**Figure 7. Dependence of distinct peaks of spider robot leg vibrations on prey robot location. (a, b)** Frequencies of distinct peaks of the vibrations of spider robot leg joints and web as a function of prey robot location on the web, for a static prey robot (a) or prey robot that moves once (b). Frequency of peak without prey robot was shown for comparison in (a). Detailed FFT's used to create this plot can be found in Fig. 6 from the accelerometers.

Thus, the spider robot at the center of the web could use the frequency of the added vibration peak after a dynamic crouch to perceive the distance of the prey robot to the center of the web, whether the prey robot was passive or moved once.

### 3. Web vibrations reveal physics behind two frequencies from system decoupling

To further confirm the origins of each frequency, we analyzed the web vibrations in the transverse direction (perpendicular to the web plane). The different spatial profiles of the amplitudes of web vibrations between the two aforementioned frequencies after a dynamic crouch (Fig. 7) further elucidated that they are most likely the natural frequencies of two partially decoupled parts of the spider robot–web–prey robot system: the higher frequency was the natural frequency of the prey robot passively vibrating on the spiral thread it was attached to, and the lower frequency was the natural frequency of the spider robot passively vibrating with the rest of the web and the prey robot.

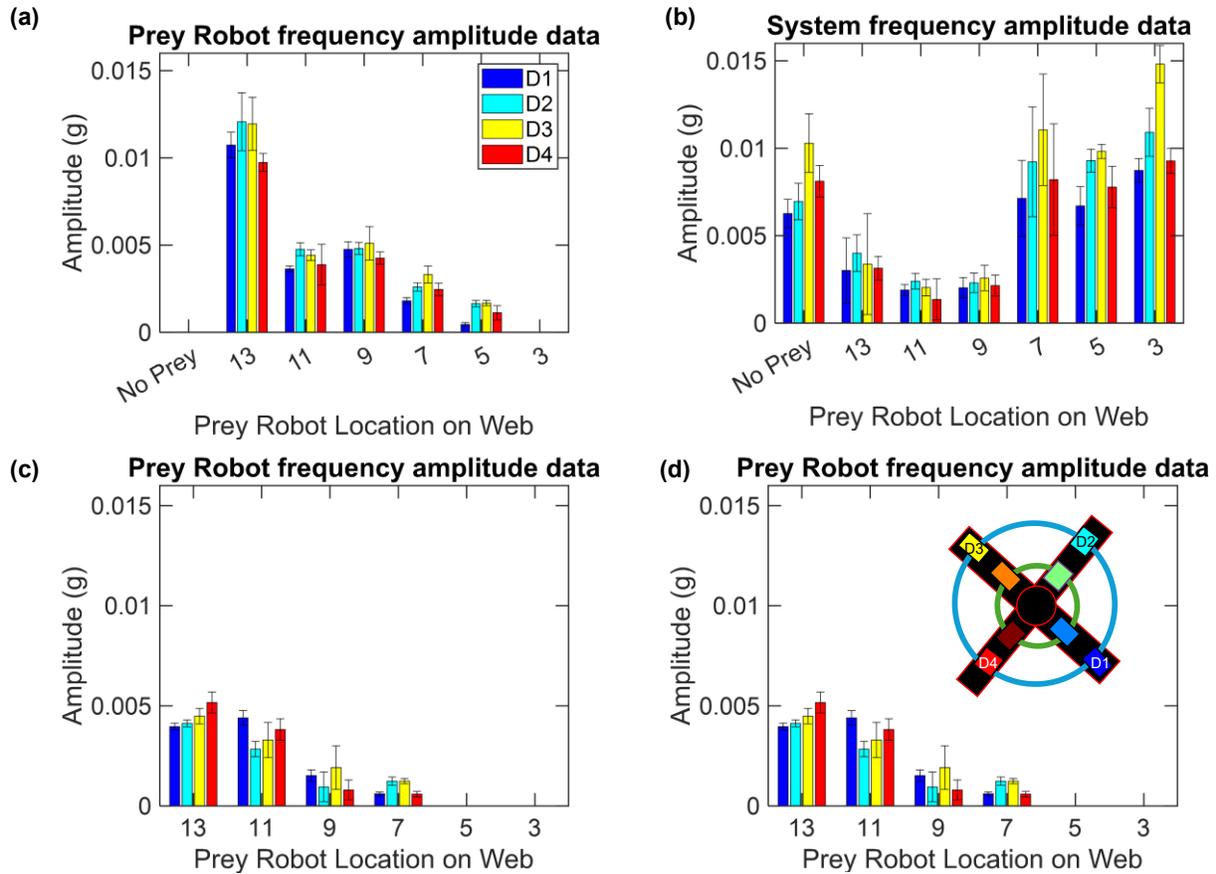

**Figure 8. Accelerometer magnitudes for each treatment (a, b, c, d)** Amplitudes of the lower spider robot–web system frequency and the higher prey robot frequency vibration peaks as a function of prey robot location on the web, for the distal joints of all four legs after spider robot dynamically crouches once when prey robot was (a, b) stationary and (c, d) actively shakes once.

Overall, the web vibrations had similar frequency components, including two distinct frequencies (Fig. 7(a), (b), dashed blue and red) that were very similar to those as observed in the spider robot leg vibrations (Fig.7(a), (b), solid blue and red whether the prey robot was static (Fig. 7(a)) or moved once (Fig. 7(b)), and regardless of the distance of the prey robot when the spider robot dynamically crouched. In addition, how the prey robot's distance affected the amplitude of each of these two distinct frequencies in the frequency domain was similar for the web vibrations and the leg vibrations (Fig. 8(c), (d) vs. Fig. 10). Specifically, the closer the prey robot was to the center of the web, the smaller the amplitude of the added, higher frequency peak was, for both the web vibrations (across all the spiral threads 1–9 tracked, except the spiral thread to which the prey robot was attached to, Fig. 10(b)-(g), blue) and the spider robot leg vibrations (Fig. 5D). In contrast, the amplitude of the lower frequency peak first decreased slightly and then increased substantially as the prey robot became closer to the center of the web, for both the web vibrations (Fig. 10(b)-(g), red) and the spider robot leg vibrations (Fig. 5C). The relative change in these amplitudes from when the prey robot was absent to when it was present was also similar between the web vibrations (Fig. 7A vs. 7B–H) and the spider robot leg vibrations (Fig. 5C, D). These similarities showed that the two distinct frequencies in the leg vibrations sensed by the spider robot were associated with the web vibrations at these two frequencies.

This shows that the frequencies of web vibrations overall matched with those of the spider robot leg joints. Even if spider joint vibration was difficult to measure when the spider was moving, the web vibration generally reflects the vibration received sensed in the joints. This was physical confirmation that the spider's senses generally reflect the frequencies of the vibrations in the web, but not necessarily the magnitude and its ratios.

How the spiral thread to which the prey robot was attached vibrated differently from other spiral threads further supported our speculation that these two peak frequencies were the two aforementioned natural frequencies. The vibration amplitude was often larger for the spiral thread on which the prey robot was attached than other spiral threads, for both the lower (Fig. 10 (d)–(f), red, highlighted) and the higher (Fig. 10(d), (e), blue, highlighted) frequencies. The latter higher amplitude meant that the spiral thread attached to the prey robot vibrated more strongly with the prey robot than the other spiral threads, showing that this thread and the prey robot were partially decoupled from the rest of the web and the spider robot. The former higher amplitude meant that the stronger vibrations at the higher frequency also made the lower frequency vibrations stronger on this spiral thread. This showed that, despite the decoupling, the prey robot's vibration still affected the vibrations of the rest of the web and the spider robot.

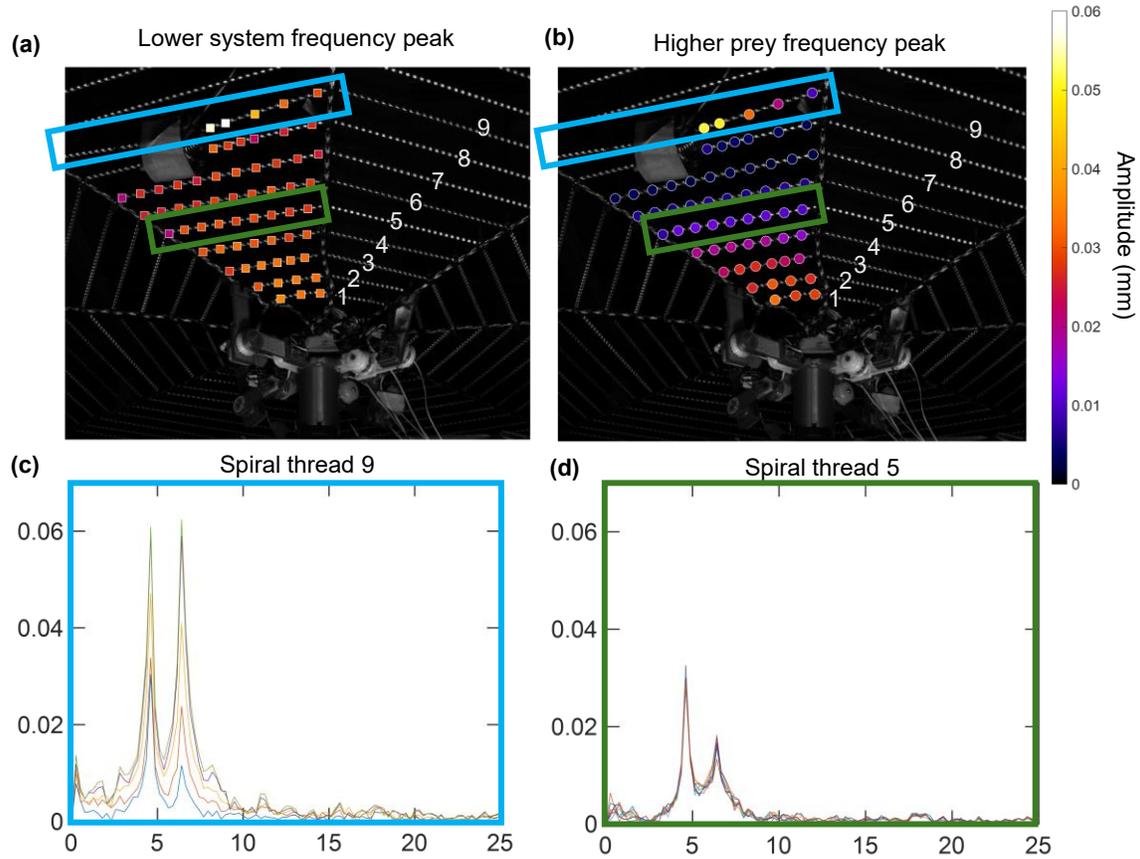

**Figure 9. Example web vibrations after a dynamic crouch.** Representative data shown for static prey robot on spiral thread 9. **(a, b)** Amplitudes of vibrations at the lower, system frequency (a) and added, higher (B) frequency peaks across all tracked points of spiral threads in a sector of the web, which is closest to the D4 accelerometer that we focus on. **(c, d)** Frequency vs. amplitude from Fourier transform of transverse web vibration (mm) vs. time, for the spiral thread 9 (highlighted blue) to which the prey robot was attached (C) and a representative spiral thread (highlighted green, spiral thread 5) it was not attached to (D). Each curve shows data from one tracked point, and data from all tracked points on the spiral thread across 3 trials are shown.

Thus, the partially decoupled prey robot contributed to both observed frequency peaks. First, the prey robot vibrating on its own spiral thread separately added a frequency peak at its natural frequency.

Second, the prey robot, via its own spiral thread's connection, effectively added mass to the system consisting of the rest of the web and the heavier spider robot, further lowering the lower natural frequency of this system.

This insight further explained why the distance of the prey robot from the center of the web was encoded in the added frequency, because the further the prey robot was, the longer its spiral thread, and the lower this natural frequency. This relationship between the prey frequency, the length of the spiral thread, and the distance of that spiral thread to the center of the web (where the spider typically lies before prey capture) takes advantage of the structure of the spiral shape in orb webs, which could explain another benefit to this structure. This geometric feature allows the spider to encode prey distance to prey frequency.

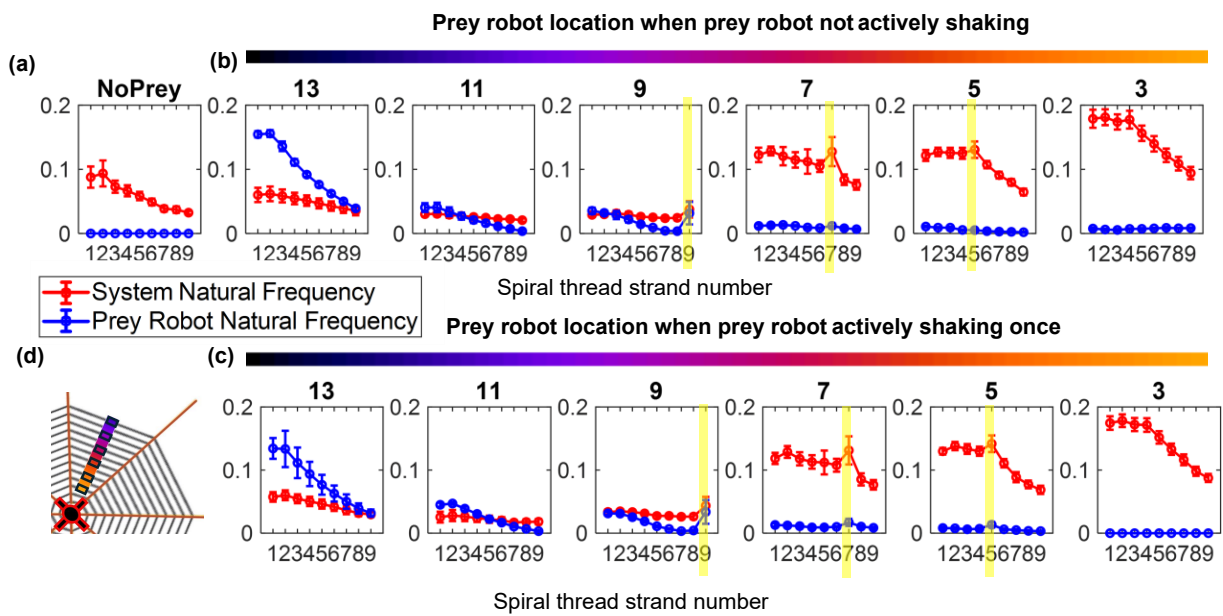

**Figure 10. Amplitudes of web vibrations at the two peak frequencies.** Data shown are for vibration amplitudes in the transverse direction (parallel to gravity), from experiments with the spider robot jerky crouching once. **(a)–(c)** Vibration amplitude across spiral threads 1–9 for the two peak frequencies (red and blue) when there is: (a) no prey robot on the web, (b) a prey robot not actively shaking, (c) a prey robot actively shaking once. Each data point shows mean ± s.d. of peak magnitudes across all points tracked across 3 trials (see representative data in Fig. 8(c), (d)). Yellow band highlights the spiral threads to which the prey robot was attached. **(d)** A color map to detail location of prey robot.

### 4. Simplified spider robot cannot yet detect prey robot orientation

Another important aspect of spider vibration sensing was identifying the direction of the prey robot. In the parallel biological study [56], the spider always identified the direction of the prey before moving towards it. However, in our system we were unable to identify how directional sensing works. The spider robot only has 4 legs, which limits the ability to model directional sensing capabilities of a spider. In all conditions when the prey robot was on the web, using the existing vibration data was not sufficient for capturing the direction of the prey robot, either through amplitude comparisons or time of arrival comparisons.

The amplitudes of the spider robot's natural frequency and prey robot's natural frequency are similar across all distal accelerometers despite them all being varying distances from the prey robot (Fig. 8). In certain cases, when the spider robot was dynamically crouching, Accelerometer D3 senses the largest vibration, despite Accelerometer D2 being closer to the prey robot. The attachment of the spider robot, or the slight variances in web tension preventing symmetry that are causing the variations in sensed parameters, and by comparing magnitude it was difficult to identify the location of the prey robot of the spider. This was supported by the observation that even if there was no prey on the web all legs experience differing magnitudes (Appendix, Fig. 17). There seems to be somewhat of a magnitude difference, but from the 4 legs there was not an easily identifiable trend.

## V. TEMPLATE MODEL

Based on these robophysical modeling insights, we developed a template model to capture the partially decoupled vibrations observed in our system. A template model is the simplest model that captures the most fundamental dynamics of a system [82]. It neglects much of the complexity of real biological or engineering systems, but in doing so provides fundamental insights that can often be generalized across systems [83–92].

## A. Models For Prey And System Frequency

Our template model reduced the complex, three-dimensional, spider robot–web–prey robot system into two partially decoupled, two-dimensional subsystems, each consisting of point masses on a thread that is attached to anchor points. For each system, we use a simple vibrating string equation to find the frequency of each system, generating the two observed frequency peaks (Fig. 11). This is very similar to a mass vibrating on a spring, where both the mass of the objects on the string and the length of the string dictate the natural frequency of the system.

### 1. System frequency

In the first subsystem representing the spider–web–prey system frequency ($f_{sys}$, Fig. 11(a)), the spider robot was simplified as a point mass $M$, attached to a long thread with length $L$, which represented the rest of the web that vibrated together with the spider robot and the prey robot. The anchors at both ends of the long thread represented the surrounding environment that the web was attached to, which were fixed anchors in our study. Because the much stiffer radial threads (rather than the much less stiff spiral threads) provide the main structural support in our physical web and in spider webs [93], they dominate the vibrations of the web. In addition, because the physical web in our study and the webs of *U. diversus* had circular symmetry, each radial thread should share the structural support roughly equally. Considering this, we used the long thread with length $L$ to represent all the radial threads on each side of the spider robot mass $M$. In addition, another, smaller point mass $m$ within the long thread represented the additional effect of the partially decoupled prey robot with its spiral thread (the second system) that contributes to the frequency of the first system.

### 2. Prey robot–spiral thread frequency

In the second subsystem representing the prey robot frequency ($f_{prey}$, Fig. 11(b)), the prey robot was simplified as a point mass m, attached to a short thread with length $l$, which represented the spiral thread that the prey robot was attached to. The anchors at both ends of the short thread represented the two much stiffer radial threads that this spiral thread was attached to. Through the spiral geometry of the web, the short thread length $l$ is correlated with the distance of the spiral thread to the center of the web.

3. **Equations for both systems**

To calculate the natural frequencies of both subsystems, we used the frequency equation:

$$f = v / \lambda \quad (1)$$

Where $v$ is the wave speed of the transverse wave (perpendicular to the web plane) dependent on:

$$v = (Tension / \mu)^{0.5} \quad (2)$$

and $\lambda$ is the fundamental wavelength and $\mu$ represents the mass per unit length. Here, we assumed that both the spider–web–prey subsystem and the prey robot subsystem only vibrate at their respective fundamental frequency on their respective threads (and not its harmonics) due to the low frequencies observed in our results, in addition to both masses, thus:

$$\lambda = 2 \cdot Length \quad (3)$$

Combining equations (1)–(3), we obtained the final frequency equation relating both the length of the string and the total mass on the string with tension, where Mass represents the mass of the total objects on the web, from $\mu \cdot L$, and assuming that the silk mass is negligible compared to the weight of the spider robot and/or prey robot, we obtained:

$$f = 0.5 \, (T / (Mass \cdot Length))^{0.5} \quad (4)$$

Once we plugged this equation into both subsystems for equation (4), we obtained the following two equations, one for the system frequency, the other for the prey robot frequency:

$$f_{sys} = 0.5 \, (T_{sys} / ((M + m) \cdot L))^{0.5} \quad (5)$$

$$f_{prey} = 0.5 \, (T_{prey} / (m \cdot l))^{0.5} \quad (6)$$

where *M* is the mass of the spider robot, *m* is the mass of the prey robot, and *L* is the length of the long radial thread equivalent, and *l* is the length of the short spiral thread the prey robot is caught on. A very similar equation of the web natural frequency has been calculated in a previous study modeling the resonant frequency of spider webs [33], but it did not consider the additional prey natural frequency.

## B. Evidence of our template model working in our robophysical model

In addition to our template model matching with qualitative results from our robophysical model, we find evidence of quantitative results matching with our robophysical experiment results.

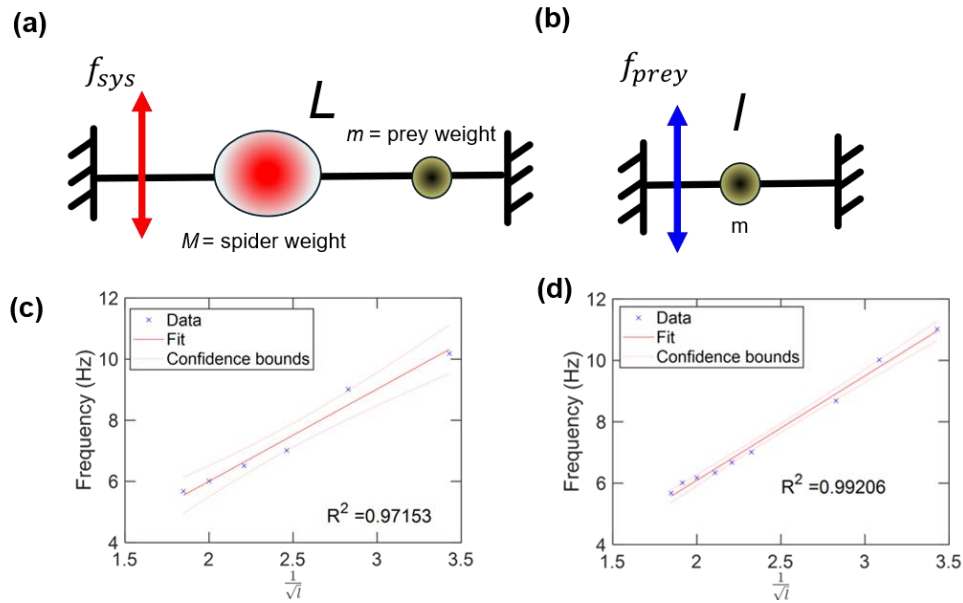

**Figure 11.** A picture of the proposed anchor model. **(a)** Anchor model for the system frequency. The frequency of the system is determined by *L* (equivalent length of web), along with *M* (mass of spider) and *m* (mass of prey robot). **(b)** Anchor model for the prey robot frequency. The frequency of the system is determined by *l* (length of spiral thread prey robot is on), along with *m* (mass of prey). **(c)** A scatter plot with a least squares regression line plotted to show the linear correlation between frequency and inverse square root of *l* from results in Fig. 6. **(d)** A scatter plot with a least squares regression line plotted to show the linear correlation between frequency and inverse square root of *l* from results in Appendix, Fig. 19. Details of calculation of Fig.11(c) and (d) are shown in Appendix, Table 1.

When there is no prey robot on the web ($m = 0$), the system natural frequency becomes larger, which reflects the results we see in our experiments (Fig. 6 and Appendix, Fig. 19). In addition, because the spider robot always stayed at the same location, its system natural frequency should not change substantially, however far the prey robot was from the center of the web, which is consistent with our model equations, as distance of the prey robot is not a variable in these equations.

This template model can predict prey robot frequency with knowledge of the length of the spiral thread the prey robot was on. Assuming that tension in all spiral threads is constant, $f_{prey}$ should be correlated with the inverse square root of $l$. Such a trend can already be seen in our interpolation of the prey robot and system frequencies (Fig. 6 and 7). We assumed that tension can be constant across spiral threads since they are not stretched but rather placed on the radial threads. In our main experiment, the prey robot frequency and the inverse square root of $l$ had a linear relationship that fit the data excellently with a correlation coefficient of 0.97 (Fig. 11(c)). An additional experiment (Appendix, Fig. 19) that varied the location of the prey robot on all spiral threads also showed a high correlation coefficient of 0.99 (Fig. 11(d)).

These results suggested that this template model is a reasonable (though much simplified) representation of our robophysical model and that the second higher frequency is the prey robot's natural frequency. The high correlation suggests that our original assumption that tension is constant across spiral threads was also sound. This also confirmed our speculation that the small, prey robot frequency peak that we originally were not sure of in spiral thread 3 due to a small amplitude that was comparable to noise (Section IV.C.1, Fig 6(b) and 6(c)) was noise.

## C. Implications of template model

The template model also suggested that, by knowing the system frequency when there was no prey robot, the spider robot may be able to infer the weight (heavier or lighter) of the prey robot by detecting how much the lower system frequency changes. The template model predicts that, when the mass of the prey robot increases, the system frequency decreases. Although prey robot frequency also changes with prey robot mass, prey robot frequency is also dependent on the length of the spiral thread, so only utilizing the prey robot frequency could be difficult to determine the prey robot weight. By identifying the system frequency, the spider robot can potentially estimate the weight of the cause of the higher prey robot frequency. This could be a potential way for spiders to characterize different objects on the web, such as mate, offspring, and debris, as these objects produce different frequency components [27,32]. By knowing the weight of the prey robot through the system frequency, the spider can potentially know the distance of the prey robot through the prey robot frequency, despite being dependent on both distance and mass of the prey robot.

This model suggested the importance of the spiral thread that the prey is on as a fundamental part of the sensing process. In other web modeling papers, the web was approximated as a continuous membrane [60,94–97] or treated as a whole structure [33]. This led to assumptions that the web was one entire resonant structure with global eigenmodes, determined by the stiff radial threads responsible for vibration transmission and structural support. Thus, these previous studies have primarily investigated the radial threads' relevance in vibration sensing. Our system frequency derivation is inspired by this concept. In fact, a very similar equation of the system natural frequency has been calculated in a previous study modeling the resonant frequency of spider webs [33]. While these models described the web's overall dynamics, our work extended this by modeling the local interaction of the prey (robot) and its corresponding spiral thread. Our template model supported by the robophysical experiments pointed to an important role of the spiral thread, not just in capturing the prey, but also in the vibration sensing process.

### D. Limitations of template model

While our template model could accurately predict the prey robot frequency based on the spiral thread length, it did not consider the effects of the locations of the spider robot and prey robot on the web, which likely affect both the system and prey robot frequency. Our robophysical model showed that $f_{sys}$ changes slightly as the prey robot distance changed. The model assumed that the mass along the spider–web-prey–system was distributed evenly, when it was more akin to two masses (the prey robot and spider robot's) on a vibrating string. These two masses will change the system frequency based on where they are located along the string, which the model did not account for. However, this prey robot location did not significantly change the system frequency (Fig. 7).

## VI.　EVIDENCE FROM ANIMAL DATA

Using robophysical model, we discovered a physical mechanism that can be used by spiders to detect prey presence and location. To further validate the biological relevance of this physical mechanism, we analyzed video recordings of fly capture events by *U. diversus* from the parallel biological study [56].

In this section, we: (A) describe the biological data used; (B) describe the vibration analysis of mentioned biological data, and analyze the mentioned biological data, by (C) analyzing jerky crouch events, and (D) analyzing other events that produce the two-peak-frequency physical mechanism.

### A.　Biological data used

In the parallel biological study, adult *U. diversus* females built a horizontal web in a 10 cm × 10 cm circular arena illuminated with a ring of white light LEDs. Web vibrations were recorded with a high-speed camera on top of the arena with a 16 mm fixed-focal-length lens at 1000 frames per second (1280 × 1024 pixel resolution). A side camera (1440 × 1080 pixel resolution) simultaneously recorded spider leg movement at 100 frames per second with a 12 mm fixed-focal-length lens. Both cameras were synchronized via a trigger signal. A single *Drosophila melanogaster* was dropped on the web to measure the web and prey vibrations as the spider captured the prey. See the parallel biological study [56] for more detailed experimental protocols.

From our robophysical modeling and template model, we identified one of the frequencies as the system frequency, which is the natural frequency of the physical web, spider robot, and the prey robot passively vibrating together. This is not a new concept, as mentioned in Section IV.C.3, and also observed in other spider webs [20,98], but the physical mechanism was not clear. In the parallel biological study and our observations of tracked videos from that study, a continually reoccurring 10 Hz frequency peak within the vibration spectrum is present every time the spider perturbed the web substantially [56]. Furthermore, the parallel biological study performed a control experiment replacing the spider with a 3-D printed object with a mass matching that of the spider (with no captured prey), and when that object was touched and released, there was the same 10 Hz signal, confirming that this frequency corresponds to the web's natural resonance [56]. Thus, this 10 Hz vibration frequency is analogous to the natural frequency of the spider–web–prey system in our robophysical model.

In addition, the consistent ~10 Hz system frequency in the spider web vibration data can be potentially explained by the fact that all the spider webs were built on the same 10 cm × 10 cm frame, with a comparable density of silk, and that *U. diversus* (0.0016 ± 0.0002 g, mean ± SEM; $n$ = 24 individuals) *D. melanogaster* flies (0.0160 ± 0.0015 g, mean ± SEM; n = 48) typically were very similar in weight between each experiment [56]. Even if the web was slightly different in shape or geometry, due to the consistent size of the webs and weight of the spider and prey, this system frequency of 10 Hz likely changed little, as the biological study found [56].

In our main robophysical modeling experiments (Section IV.C), we found that a jerky crouch produces post-crouch vibrations in the web with two distinct frequency peaks that *can* be used by the spider robot to identify prey robot presence and location, with the lower frequency corresponding to the system frequency. Thus, if such a physical mechanism occurs in the biological system, its web vibrations should include the 10 Hz natural frequency of the biological system as just described. Therefore, for the analysis here, we chose the animal trials from the parallel biological study in which we could identify web vibrations with the 10 Hz component using the same centroid-based tracking points utilized in our physical web (Section II.C). The parallel biological study focused on the broader behavioral transitions of the spider in close loop with prey behavior on the web, rather than the role of jerky crouches here, and thus jerky crouches were not observed in all the videos. In addition, our analysis here required tracking specific points on the web (see Section VI.B), whereas the biological study used a pixel-intensity fluctuations to measure whole web vibrations continuously along all threads [56]. We also required a side view video with visible movement from the parallel biological study, which mainly used top views for vibration analysis and side view videos for broad quantification of behavioral transitions of the spider and prey. As such, there were only 10 trials of prey capture videos in which there were both top and side views of the spider.

Out of the 10 trials of prey capture videos, two met the 3 requirements: the trial had a spider producing a distinct jerky crouch motion, both the fly and the spider are clearly visible in both side and top view videos, and the spider produced the large 10 Hz frequency peak vibration in the web. We analyzed the vibrations during and after the spiders' single jerky crouch of these two trials. The large 10 Hz frequency peak was also observed in locomotion behaviors and were analyzed later (Section VI.D).

In these two trials, if we could find similar passive vibrations occurring with a second, higher frequency component, which could potentially be the natural frequency of the prey on its spiral thread, then the physical mechanism that our robophysical modeling revealed is likely one of the strategies used by the spider (see Section VII.D for discussion of future work that is needed to confirm this).

**B. Vibration analysis of biological data**

Similar to the spider robot leg vibration analysis (Section II.B) and the physical web vibration analysis (Section II.C), an FFT was performed on the vibration events on the top videos (because we focused on analyzing web vibrations to see if there is an additional frequency higher than the 10 Hz frequency, as explained in the section above). Like our analysis of the spider robot leg, we temporally cropped the vibration data to only include the single jerky crouch behavior that we are interested in here. This was to ensure that the analyzed vibration was not too noisy, as many different frequencies occur in the web throughout an entire trial due to many other behaviors occurring closely before or after jerky crouch (e.g., static, shaking, soft crouching), which cause the FFT results to be very noisy. Naturally, real spider behaviors are much more diverse and less distinct than the very well controlled and repeatable robot jerky crouch that we used in this first robophysical modeling.

Events were classified by observing the side videos (higher resolution of spider) using our classification of short-timescale leg behaviors (Fig. 1) and looked at locomotion behaviors (walking, turning), both of which produced 10 Hz resonant vibrations and temporally cropped the vibration data, from when the web began to produce this large 10 Hz resonant frequency component, to when this frequency component disappeared. We classified this time frame as an event.

For the chosen animal trials, we measured the web vibrations via the top view high-speed camera (1000 frame/s), mainly because it had the higher frame rate (10 times faster). Due to the spider moving not only its legs but also its entire body, it was very difficult to accurately track the spider's vibrations (as the spider constantly rotated, involving yaw, pitch, and roll, which 2-D tracking cannot track), so only the prey and parts of the web that could be tracked (bright reflections at the intersection of threads of the web) were used in this analysis. We used dltdv8a [79] to perform dot centroid-based tracking on the points of interest (prey and select points on the web) from the top view. A rectangular window was used before applying the FFT. Although this resulted in spectral leakage at ~1 Hz (due to the discontinuity of start and end values), it helped preserve the frequency resolution given the low sample size (due to the short duration of the vibration, ~1 seconds or ~1000 frames) for each analyzed event. This ~1 Hz component can be neglected, as such low frequency motions were not observed in the raw video data.

Most of our FFT methodology was similar to the spider robot leg analysis (Section II.B) and the physical web vibration analysis (Section II.C). The difference was that, in the robophysical modeling data, we had to crop out any spider robot movement due to the large jerky crouch actuation that produced the 1 Hz frequency in the FFT; otherwise, it obscured any other frequencies detected. But in the animal data, the spider does not produce as large a motion, so there was no such effect on the FFT. We instead analyzed the entirety of the event so that the initial vibration of the spider movement was also included in the FFT analysis. This was done because the jerky crouch of the spider did not cause the sharp peak that was easy to identify and isolate for analysis as done in the robophysical modeling (Section. II.B; Fig.3(d)) (this was likely because the spider robot and physical web were heavier compared to the physical web tension than the spider and its web compared to their web tension; see Section VII.D for discussion of this limitation).

## C. Identifying the 2nd frequency mechanism from jerky crouch

From the video analyses, we found that in the two events where the spider caused the web to produce vibrations with the ~10 Hz frequency component, through a jerky crouch, it also caused the prey to oscillate at a second frequency. Both the tracked web points and prey showed a lower frequency component, typically around 10 Hz (Fig. 11(b) and 11(e)). However, the tracked prey also showed an additional frequency component at a higher frequency of 10–25 Hz (Fig 11(b) and 11(e)).

This observation can be made not only by analyzing the FFTs, but also from the raw vibration data. Comparing to an ideal 10 Hz only exponentially decreasing sinusoidal wave (Fig. 11(c)) and to a 10 Hz exponentially decreasing sinusoidal wave with an additional small 12 Hz component (Fig. 11(c)), the latter shows distortion (the wave briefly oscillates at a low amplitude compared to the beginning and end) and resembles the observed vibrations in Fig. 11(e). In addition, the 10 Hz oscillation of the prey was slightly out of phase with the rest of the tracked web points, providing further evidence that the prey was shaking differently from the rest of the web and the spider.

While prey do struggle on the web, they freeze during strong bouts of spider movement [56]. During the short period after a jerky crouch, the fly is still in this frozen state and moves passively as a result of the web vibrations produced by the spider. The point with the strongest second frequency vibration was the prey itself in both trials. This suggested that the second, higher frequency peak originated from the prey.

Comparing these spider web vibrations (Fig. 11(b) and 11(e)) with our physical web vibrations in our robophysical modeling main experiments (Section IV.C.3), both observed that different portions of the web vibrate differently, in a similar fashion. Because the spider only crouched when there was a disturbance on the web, it was hard to confirm whether the two frequencies would become one when the spider was alone on its web, similar to what was seen in our robophysical model (Fig 6(a)). However, as seen by the prey vibration (Fig 11), there was a second peak frequency from the prey, in accordance with our robophysical modeling results (Fig 6(b) and 6(c)). This indicated that our model reflects certain prey capture events.

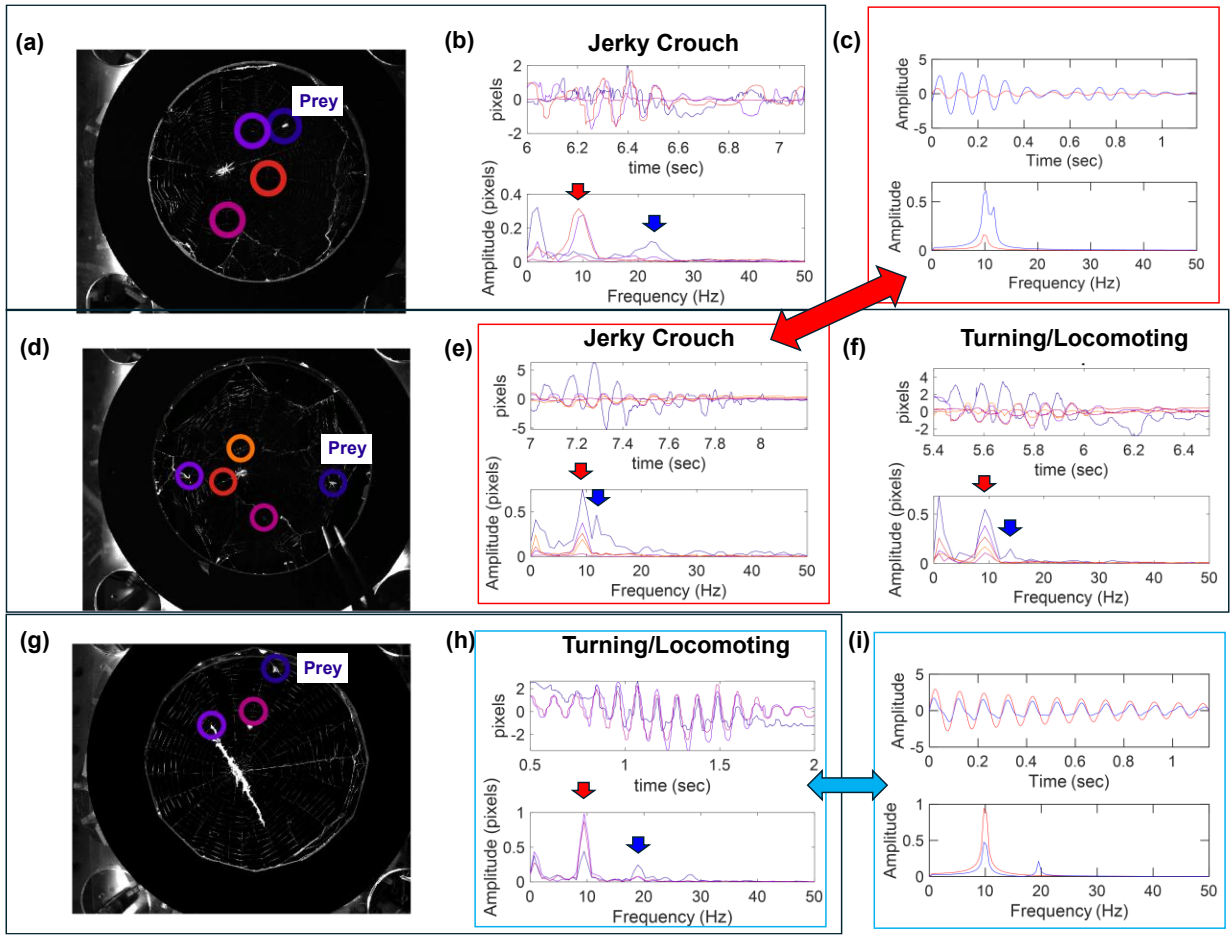

**Figure 12. Vibration analysis of fly and stabilimentum during three prey capture videos.** The tracked videos are on the left **(a, d, g)** and correspond to the extracted raw data (top) and resulting FFT (bottom) was presented for each treatment within a box. All events present the first frequency at around 10 Hz (seen in red arrow). The blue arrow points to the frequency where prey is shaking at. The prey is circled in dark purple. **(b, e)** Sample treatment for jerky crouch, where the prey is additionally vibrating at (b) ~23 Hz or (e) ~12 Hz. Video example of (e) can be found at Movie S6 [165]. **(h, f)** Sample treatment for turning/locomotion, where the prey is additionally oscillating at (h) ~19 Hz or (f) ~14 Hz. **(c, i)** A sample

recreation of first frequency (red) and second frequency (blue), (c) of event analyzed in (e), and (i) of event analyzed in (h).

D. **Identifying two frequencies in locomotion events**

As mentioned in Section VI.A, locomotion (turning and walking) also induced the 10 Hz frequency vibrations. Using the same analysis methods as in Section VI.B, we analyzed 2 locomotion events from 2 trials and found that the second frequency peak also appeared during these events when prey is on the web. (Fig. 11(f), and 11(h)). Again, comparing to an ideal 10 Hz only exponentially decreasing sinusoidal wave (Fig. 11(h)) to a 10 Hz exponentially decreasing sinusoidal wave with an additional small 20 Hz component (Fig. 11(i)), the latter showed distortion (the wave never fully dips down compared to the former vibration) and resembled the vibrations observed in Fig. 11(h). This confirmed that the second peak frequency did exist in the raw vibrational data.

These results indicated that jerky crouching, walking, and turning all may sufficiently vibrate the web enough to induce the prey to shake at its natural frequency on its spiral thread. Indeed, upon closer visual inspection of these events of the videos we analyzed, the prey vibrated slightly differently than the rest of the web during these events (Movie S6 [165]). This was further evidence that the physical mechanism we discovered using our robophysical modeling occurs within the biological system. These analyses showed that the physical mechanism *exists* within the system and that it *can* be used as a strategy by our model organism and perhaps orb weavers in general to locate and detect prey.

This analysis of the animal data was a good example of "robotics-inspired biology", using robots as active physical models to study biological systems to generate testable biological hypotheses [66,67,70–72] . We would not have been able to identify this mechanism without the systematic study of our system beforehand, as an uninformed analysis of the videos would not have revealed this second frequency as it is easily obscured by noise.

## VII. DISCUSSION

### A. Summary of findings

We created a robophysical model consisting of a spider robot, a physical web, and a prey robot for systematically studying orb-weaving spiders' use of short-timescale leg behaviors for vibration sensing of prey on a web. Our robophysical model enabled a repeatable experimental setup, a controllable spider and prey model, and simultaneous multi-modal data measurement, which are not practical in biological experiments.

We discovered that a jerky crouch by the spider robot causes the entire web and itself to vibrate passively afterwards at the low natural frequency of this system. This vibration further induces the prey robot to passively vibrate on the web, at not only this lower frequency but also an added, higher frequency. This added, higher frequency was the natural frequency of the prey robot on its spiral thread. The closer the prey robot was to the center of the web, the higher this frequency was. This was because the spiral geometry of the web causes the length of this spiral thread to correlate with distance. Thus, the spider robot can detect the presence of a prey robot by detecting the added vibration frequency and sense its distance by sensing this frequency utilizing the geometry of the model web. We also found evidence that this occurred in the animal data of the parallel biological study, by identifying a second frequency. Visual inspection of both robophysical experiment and animal videos showed that the prey robot or prey vibrated passively at a higher frequency than the rest of the web and the spider robot or spider (Movie S7 [165]).

To further explain our results, we developed a template model to capture the most fundamental features of the dynamics of the two decoupled systems: the prey robot vibrating on its spiral thread, and the spider robot and the rest of the web. Our template model further quantitatively predicted how the system frequency changed depending on the equivalent length of the web and the spider and prey weight, and the length of the spiral thread based on what the prey natural frequency was. This model highlighted that the spiral thread with the prey was decoupled from the web and a critical part of the prey vibration sensing process, an additional feature to build on previous models and observations. Building upon previous work that focused more on the role of radial thread vibrations, our work suggested that spiral thread vibrations may also play an important role on spider vibration sensing of prey (and other objects) on the web.

Our study helps expand robophysical modeling of biology to mechanosensing in complex environments (i.e., vibration sensing on a web). While robophysical modeling has been widely used for studying biological locomotion (e.g., [99–134] (for reviews, see [66,67,70–72,135,136]), its use in biological sensing has focused more on sensing visual patterns [137,138], geometry [139,140], sound [141], flow [142], tactile [138,139,143,144], force/pressure [145–149], and electrical field [150–152]. Here we demonstrated its usefulness to circumvent the major limitations of biological studies of spider vibration sensing. Without our robophysical modeling observations, we would not have arrived at the simplified template for the highly complex biological system. Our observations allowed us to simplify it into an easily interpretable template model without losing biological relevance, providing fundamental insights not previously found through biological observations [2,26,33,57] or complex simulations [94–96]. Here our robophysical modeling added quantitative evidence for a physical mechanism that *can* be used by orb-weaving spiders to sense prey on the web.

Our findings also suggest that web spiders may use short-timescale leg behaviors as a new form of active sensing, given their unique ability to use their built environment (web) to extend their vibration sensing capability. Rather than the spider passively waiting for prey actively moving to generate cues, the spider could be using dynamic leg crouching to induce the prey to passively vibrate on the web. Our robophysical modeling showed that this would shape the sensed information in a way that may help the spider better detect prey presence and distance even if the prey is not actively moving on the web.

## B.  A proposed active sensing strategy based on our observed physical mechanism

Our findings provided quantitative evidence and a mechanistic explanation of the long-standing hypothesis that spiders dynamically crouch to detect an "echo" [2,26,27,53,55,57]. In our findings, an active signal (in the spider's case a jerky dynamic crouch inducing the system frequency) was generated and the resulting response (an "echo", in this case the induced prey-passive vibration frequency) was then used for sensing purposes. While these hypotheses resemble echolocation in bats and toothed whales [153,154] that emit vibrations (sound waves) and uses the echoes to locate prey, the physical principles due to the time-of-arrival mechanism of sound-based echolocation does not work in the web.

In bat-based echolocation, the emitted pulses are at a high frequency (>10 kHz) over a short duration (a few milliseconds) and rely on a time delay between the generated pulse and reflected echoes to interpret the location and identity of a target [153]. By contrast, as our robophysical modeling showed, the spider's dynamic crouch produces low-frequency vibrations (on the order of ~10 hertz) in the web that persist over a long duration (a few seconds), much longer than the travel time of the signal across the web. This presents a unique problem of an overlapping input and output signal that occur almost simultaneously that is not present in echolocation.

Based on our low frequencies, we can calculate that they are temporally overlapped in our model. By calculating the wave speed of transverse waves in a thread using equation (1), the wave speed in our model web was 880 m/s. The length of the radial threads of the physical web was only 40 cm, so at most there was a 0.4-millisecond lag between when the spider robot vibrated to when the prey robot vibrated, which was too short for the spider robot's own vibration signal to dissipate. Low frequencies observed both in biology studies and our robophysical model require at least a few hundred milliseconds to be interpreted as a frequency. This is due to the relatively longer wavelength of a low frequency. For example, a low frequency of 50 Hz has a period of 0.02 seconds. Before a 50 Hz frequency vibration completes one period (which is the mathematically minimal amount of time required to identify a frequency), the return signal would have arrived at the spider robot already. This does not happen in echolocation (as in bats) due to the very high frequencies it produced over the longer distances. A 10k Hz frequency wave as in bat echolocation [153] has a period of 0.1 millisecond, which can oscillate multiple times before the "echo" returns to the echolocation user (typically on the order of tens of milliseconds). Similar to this, both the prey robot and the spider robot's natural frequencies are too low to be properly separately temporally in a small sample time. Thus, in our robophysical model, the two peak frequencies must overlap even if they are from different sources.

The same explanation can be applied to the spider webs of the parallel biological study [56]. In the spider videos, the spider's emitted signal, which is the natural frequency (10 Hz) of the system passively. Given the small spatial scale of a spider web (10–100 mm) the "echo" takes less than a millisecond to propagate through the web (at 100–500 m/s) [27,155]. This is well below the period of the system frequency (e.g., a period of a 10 Hz signal is 0.1 seconds). Unlike echolocation, where the signal and echo are separate, this sensing strategy uses low-frequency vibrations that overlap. The spider must therefore interpret the emitted and "echo" signals simultaneously.

Finally, the quantitative behavioral transition analysis of the parallel biological study supported that spiders use this potential form of sensing [56]. It identified that *U. diversus* tends to pause their movements when the fly was moving, and crouch/shakes when the fly decreased its movement on the web [56], which implied that dynamic crouching was done to increase fly movement on the web, perhaps for prey detection and localization (although that study focused more on active prey behavioral response, rather than passive prey vibrations as our physical mechanism suggested to be important here).

## C. Implications for spider vibration sensory biology

We discovered a physical mechanism that *can* occur when the spider jerky crouches (or locomotes) on the web. This physical mechanism occurs when the spider vibrates the web–spider–prey system at the system's natural frequency, which induces the prey to vibrate on its spiral thread at a higher frequency.

In the past, due to the high-pass filter mechanism of spiders and the many low-frequency vibrations originating from natural sources, researchers have focused on small-amplitude, high frequencies (200–1000 Hz) in spider vibration sensing [20,23,27]. Many papers cite the potential relevance of the low-frequency vibrations due to the low frequencies produced by caught prey, but have not suggested how they may be useful to the spider [20,27,33]. This may also be partly due to the high-amplitude vibrations that low-frequency web vibrations have within the web, which cannot be fully measured by the laser doppler vibrometer commonly used to study spider web vibrations [27]. On the other hand, both our work and the parallel biological study [56] identify low-frequency, high-amplitude frequencies to be relevant. This supports previous evidence of the importance of low-frequency high-amplitude vibrations to trigger spider prey capture behavior, where larger amplitude vibrations (much larger than the high-frequency vibrations in echolocation [153]) that were required to elicit a reaction from orb-weavers [26]. This study opens the possibility that the spider may be relying on both cues from large-amplitude, high-frequency vibrations and small-amplitude, low-frequency vibrations to detect prey.

Our study is a first step of using robophysical modeling to identify the possibility that spiders actively engage in behavior that has some ethological utility. Because biological experiments have little control over how the spider and prey behave, our robophysical modeling provided a principled and controlled approach for understanding this problem. It helped uncover a property that we had initially not noticed—that this second prey resonant frequency occurred right after the spider generated large web vibrations that induced the prey to passively vibrate. It provided a physical explanation as to why spiders would use jerky crouches, supporting the idea that there may be active sensing involved.

### D. Limitations of our robophysical modeling

While our robophysical model here proved useful in revealing physical mechanisms and principles otherwise difficult to obtain, it can be improved in multiple ways for further understanding this complex biological problem and vibration sensing by spiders more broadly. Some of these improvements require innovations in engineering development, while others need further biological studies to provide guidance and constraints to make robophysical modeling more accurate.

Importantly, our robophysical modeling was carried out in parallel with the biological study of the model organism [56]. The parallel biological study investigated more broadly behavioral transitions during prey sensing and capture. It found that crouching and shaking can induce prey movement and increase vibratory power on the web but did not analyze how joint-level vibratory power changes during prey capture. This analysis requires highly precise joint annotations, which the current tracking pipeline does not yet achieve. By contrast, our robophysical model systematically investigates the vibrations within the system to the joint-level, enabling a systematic exploration of the system, that was not feasible with freely behaving spiders. However, it took substantial time to obtain and understand the biological data; thus, some of the choices we made in the robophysical modeling here were made before we fully understood well the biological model system (*U. diversus* sensing prey on its horizontal web) and relied more on the related biological literature. Conversely, key insights from the robophysical modeling that took multiple iterations of experiments and analyses to emerge were not available at the experimental stages of the biological study. These insights made it clear only later what aspects of the biological system were more crucial to quantify to inform and improve the robophysical model. This gradual iterative process highlighted the challenges of such interdisciplinary collaboration to apply the robophysical modeling approach to new biological problems. We hope that discussion of the limitations of our work in this context can help other researchers use the robophysical modeling approach more effectively.

First, the spider robot was greatly simplified, so its mechanical design and fabrication and actuation can be improved to more accurately to capture the morphology and behavior of the real spiders. This involves 4 key limitations: (1) increased leg numbers and adding bilateral symmetry, (2) increasing the number of joints, (3) increasing joint range of motion, (4) improving the actuation profile. A new spider robot could consist of 8 legs arranged in a bilateral, morphologically accurate orientation rather than the 4 legged-radially symmetrical spider robot that we had. With this, we would be able to further analyze how the spider can sense prey direction, a key feature into spider vibration sensing we have not understood yet through our model. In addition, our robot did not have the proper joint count or structure of a typical spider. With more representative joints, we can investigate how these vibrations travel up the joints of the spider's legs. Previous research has indicated that there potentially might be some benefit to having vibration sensors in multiple joints, but with our limited robot, we found out that our proximal accelerometers did not provide any interpretable signal. Our current spider robot was also limited in its actuation capabilities, barely capable of crouching. The actuation of the spider robot can be improved by replacing the current actuation method which involves strings to be attached outside of the robot so it can dynamically crouch to a tendon-driven system where the motor pulls a string inside the spider legs. This will help decrease potential artifacts of vibration resulting from the actuation mechanism being outside of the spider body.

Our team is developing a more biologically accurate spider robot [73]. Initial tests of this new, more accurate robot on our model web confirmed the same two-frequency mechanism, further validating the core findings of this paper. In the future, we plan to use this model to further validate the template model proposed and reveal how directional sensing would work that this paper could not show.

Similarly, the prey robot was greatly simplified and should be made more biologically accurate for future studies. When the prey moved on the web, it was very lightly moving. In this paper, we replaced this with a singular light impulse on the web, but in reality, the prey occasionally moves through brief struggles, which the singular light impulse does not fully represent. We need to replace this with a very weak unbalanced motor to create occasional light vibrations instead of causing a solenoid to produce a light impact on the web. If our current prey robot had tried to occasionally move by turning on and off, repeating the light impact, there would be a substantial vibration resembling continuous moving, which was uncharacteristic of the real prey vibration. Switching to a more representative vibration profile may help us identify the effects of active vibration versus passive prey vibration during the spider sensing process. In addition, in this paper we only test the prey robot shaking in the longitudinal direction of the spiral thread, but we had not tried impulses in other directions. While it probably would not affect our core results (the dynamic crouching from the spider robot overrides these small vibrations), it would affect how the spider robot senses the prey robot movement while static, which would be a potential point of interest that we do not focus on in this paper.

In addition, our physical web was also greatly simplified and physically constrained compared to real spider webs. The radial threads of our physical web were relatively tight and sagged less under the weight of the spider robot and inertial forces during dynamic crouching (Movie S4 [165]), as compared to the larger sagging of the less tight web of *U. diversus* in the biological study (Movie S7 [165]). This may affect the resulting vibrations in the real web by potentially increasing the magnitude of the vibration or increasing the two natural frequencies of interest. This was because there was no mass-produced man-made material like spider silk, with the low-density values and high Young's modulus. While we may have increased the size of the web ten times compared to the real web, the threads of our model web are 1000 times thicker. This would greatly increase the weight making it a non-trivial factor to the frequency of both the prey and system frequency according to our template model. We suspect that a combination of the increased tension and increased weight compared to the real system was probably the reason why the two frequencies that we find in the real biological system were approximately 2 times higher than the frequencies we find in our model web. Moreover, our physical web (360% of the mass of the spider robot) was relatively heavier than real spider webs (<100% of spider mass [156]). This would result in only lower frequencies being able to propagate in our robophysical system compared to the higher frequencies that other biological studies are able to measure [23,27,98]. Making artificial silk that has the exceptional mechanical properties of real spider silk yet is still lightweight is a long-standing engineering challenge [157].

To assist in improving our robophysical model's accuracy, we would need more biological measurements to compare and validate our model. These would include: (1) The 3-D kinematics of the spider to understand the motion profile of different behaviors to implement in the spider robot. (2) The 3-D motion profile while the prey is struggling to replicate prey motion better, to identify if active prey motion can also help the spider detect some information about the prey. Our examination of videos from the biological study did not measure active vibration. (3) The stiffness values of the spider legs when they are relaxed and when they are tensed up to scale stiffness in the joints appropriately. (4) The tension distribution within a web to ensure that the model web is representative of the real web. (5) The mass distribution within the spider to ensure that the spider robot is appropriately balanced.

If we can obtain these measurements in the biological system, we can better dynamically scale future robophysical models by matching the ratio between different types of physical forces so that the robophysical model and the biological system are more dynamically similar [158].

While there are limitations to our robophysical model as mentioned in this section, we have demonstrated in this paper that robophysical modeling is a powerful tool that can identify physical principles of biological systems. By improving the robophysical modeling past the limitations mentioned here, we can use this modeling approach to generate and test new biological hypotheses, like the two-frequency mechanism identified in this paper that can guide future biological experiments.

### E. Future directions

Our work demonstrated the usefulness of robophysical modeling to complement biological experiments and theoretical and computational modeling for studying spider vibration sensing. Many future directions can benefit from leveraging this approach.

First, the physical mechanisms of sensing the direction of prey (or other biotic targets such as mates or predators) on a web remain to be discovered, which could be facilitated by future robophysical modeling. Web spiders are thought to determine target direction by comparing vibration amplitudes across their legs, rather than differences in time of arrival of vibration singles across legs, which are likely too small to be detected by them [2,24,27]. The biological study showed that changes in vibration amplitude across radial threads of the web are strongly correlated with and can be used to predict the spider's orienting behavior [56], which supports this idea. However, our robophysical modeling using the much-simplified spider robot has not yet found evidence of this (across the joints, there was not much magnitude difference, Section IV.C.4, Fig. 8). This reflected a significant difference between our robophysical model with the biological system. This may come from several reasons. Our spider robot tested in this study may be too simplified to reveal this, due to its radial symmetry, only having four legs rather than eight, and substantial simplification over the real spider's legs. Our physical web's tension distribution may also be different from that in spider webs (though the only paper with spider web tension measurements has only sample measurements of radial and spiral threads and is not sufficient for assessing this [63]). We recently developed a new more biologically accurate spider robot, which has eight, more morphologically accurate legs [73] and plan to use it to test this hypothesis. If a time-of-arrival or amplitude-based strategy still does not work, future work should use a phase-based analysis method using the Hilbert Transform to properly identify the phase-lag between each of the legs to determine if the spider robot can determine direction through phase differences.

In addition, the effects of dynamic crouching when the spider is not in the center of the web are not studied in this paper. In our biological study, we found evidence that the prey frequency decreased as the spider approached the prey. Based on our template model equations, this means the spiral thread got looser as the spider got closer to the prey. Future robophysical modeling systematically varying the spider robot location may reveal how spiders continually sense prey while it moved towards prey.

Moreover, robophysical modeling can be adapted to study vibration sensing in different web architectures. Our robophysical modeling focused on the *U. diversus* spider that makes a horizontal web. In this situation, a jerky dynamic crouch by the spider causes prey to passively vibrate transversely (perpendicular to the web plane, parallel to gravity). However, most orb-weaving spiders make webs that are more vertical [159]. In that case, passive prey vibrations after a jerky crouch may have more lateral and longitudinal components (more within the web plane). In addition, we identified that the spiral shape of the orb-web contributes to encoding the prey distance with the prey frequency (i.e., embodiment [160] via the built environment for extended sensing [16]). Robophysical modeling investigation may help discover physical mechanisms for vibration sensing in these webs and reveal general principles of how the physics of vibrations of the spider–web–target system shapes these strategies. Previous studies have shown that the direction of web vibrations (e.g., longitudinal vs. transverse) is important in high-frequency sensing [2,27,98] . In our study we focused on the dominant transverse vibrations. Preliminary observations of our physical web vibrations suggested that these low frequencies were the same in all directions. Future studies should investigate whether other vibration direction components play a role in low-frequency vibration sensing on a web.

Future studies should also further investigate the physical principles behind other behaviors of the spider. In the animal videos that we analyzed (Section VI), the spiders often performed a soft crouch after a jerky crouch on the web. We also found evidence in our robophysical model that the spider robot also detected two frequency peaks when the prey robot impacted on the physical web (Section III.D.8). If this were true for the biological system, it would suggest that spiders may also be able to sense prey that impact a web without needing to dynamically crouch. In addition, future work should study other prey sensing behaviors such as jerking (different from jerky crouch, see Section I) and shaking, and even behaviors typically associated with mate communication (such as tapping and plucking, see Section I) identification of other static objects like leaves, or web-maintenance behaviors such as locating broken sections of the web during construction [161]. By changing our robophysical model to match the properties of different spiders and adjusting the actuation capabilities of the robot spider to match the behaviors that we are trying to investigate, future work can use our modeling method to investigate other behaviors and spiders. Future spider robots that combine both vibration sensing and locomotion and crouching capabilities on the web (e.g., prototype in [162]) will enable systematic studies to discover more physical mechanisms and uncover the physical principles of spider prey capture behavioral sensing strategies. Future work can also use robophysical modeling to study the effect of prey size and mass on a spider web in behavioral modulation of vibration sensing of prey, complementary to biological investigations [163].

Furthermore, future work can expand and confirm our template model in both the robophysical model and the biology. Earlier, we explained how improving our robophysical model could allow for further confirmation of our template model. However, for our model to be robust, it must be fully validated against a large amount of biological measurements. In this paper, we only provided initial evidence in the model organism that a two-frequency vibration physical mechanism *can* be used (Section VI). To fully confirm this, future work should use cameras to capture more trials that focus on vibrations of the spider, web, and prey after jerky crouches and locomotion events.

Lastly, robophysical modeling will also be useful for systematically studying closed-loop spider–prey, spider–spider (e.g., courtship [164]), or spider–predator (e.g., web invasion [39]) interactions. When spiders repeatedly pause and crouch as they approach prey, there might be an active feedback loop from the spider reacting to the active struggles of the prey. In the parallel biological study, the spider is more likely to dynamically crouch when the prey is not actively struggling on the web [56]. Similarly, web spiders can interact with different behaviors not mentioned in this paper with other agents (a potential mate, offsprings, or even a predator [39]. Future robophysical models can add feedback between the spider robot and prey robot (or other agent robot) to test how two-way communication can be done between the spider and other agent.

**Acknowledgements:** We thank Nicholas Llaruado, Kyle Taylor, and Milla Ivanova for helping with developing the robophysical model; Tyson Hedrick for providing the custom multitrack code for dltdv8a to track the web points; and Beth Mortimer, Sunny Jung, and Noah Cowan for discussion.

**Funding:**

This work was supported by an NSF Physics of Living Systems grant (PHY-2310707), co-sponsored by NSF Dynamics, Control and Systems Diagnostics to C.L. and A.G., a Burroughs Wellcome Fund Career Award at the Scientific Interface to C.L., and an NIH grant (R35GM124883) to A.G. H.H. discloses support from The Ministry of Education (MOE) Taiwan Scholarship Program.

**Author contributions:**

C.L. and A.G. conceptualized study and acquired funding. C.L. supervised study. E.H.L., C.L., Y.Z., and L.M. developed methodology. E.H.L conducted investigation, formal analysis, and data visualization. H.H. and A.G. provided animal data analyzed. E.H.L. and C.L. wrote the manuscript with feedback from H.H. and A.G.

**Competing interests:** Authors declare they have no competing interests.

**Data and materials availability:** The data that support the findings of this article are openly available.

**Supplemental Material:** This supplemental material provides additional information that complements the main results presented in the article. It includes a list of Supplemental Movies of the behaviors of the spider and the spider robot described in this paper.

**Appendix**

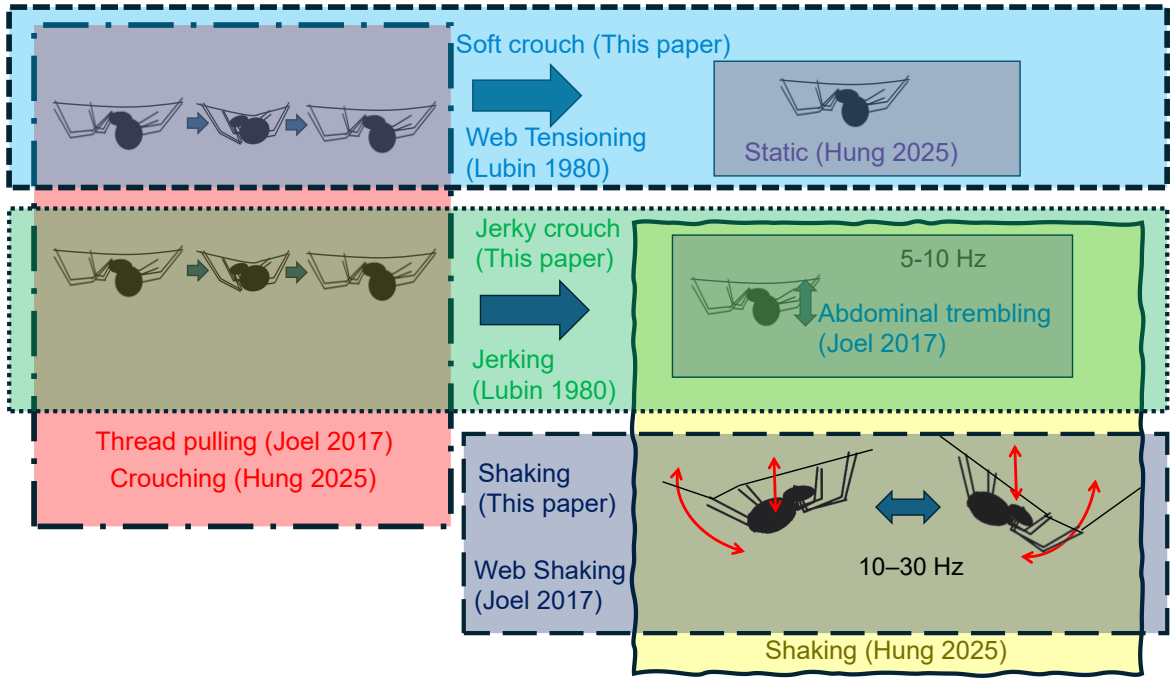

**Figure 13. A comparison of different terminology for active spider behaviors during prey capture.** The terminology is from a combination of three papers [50,51,56] and this paper (Section I). In this paper, we use soft crouch, jerky crouch, and shaking to describe these behaviors.

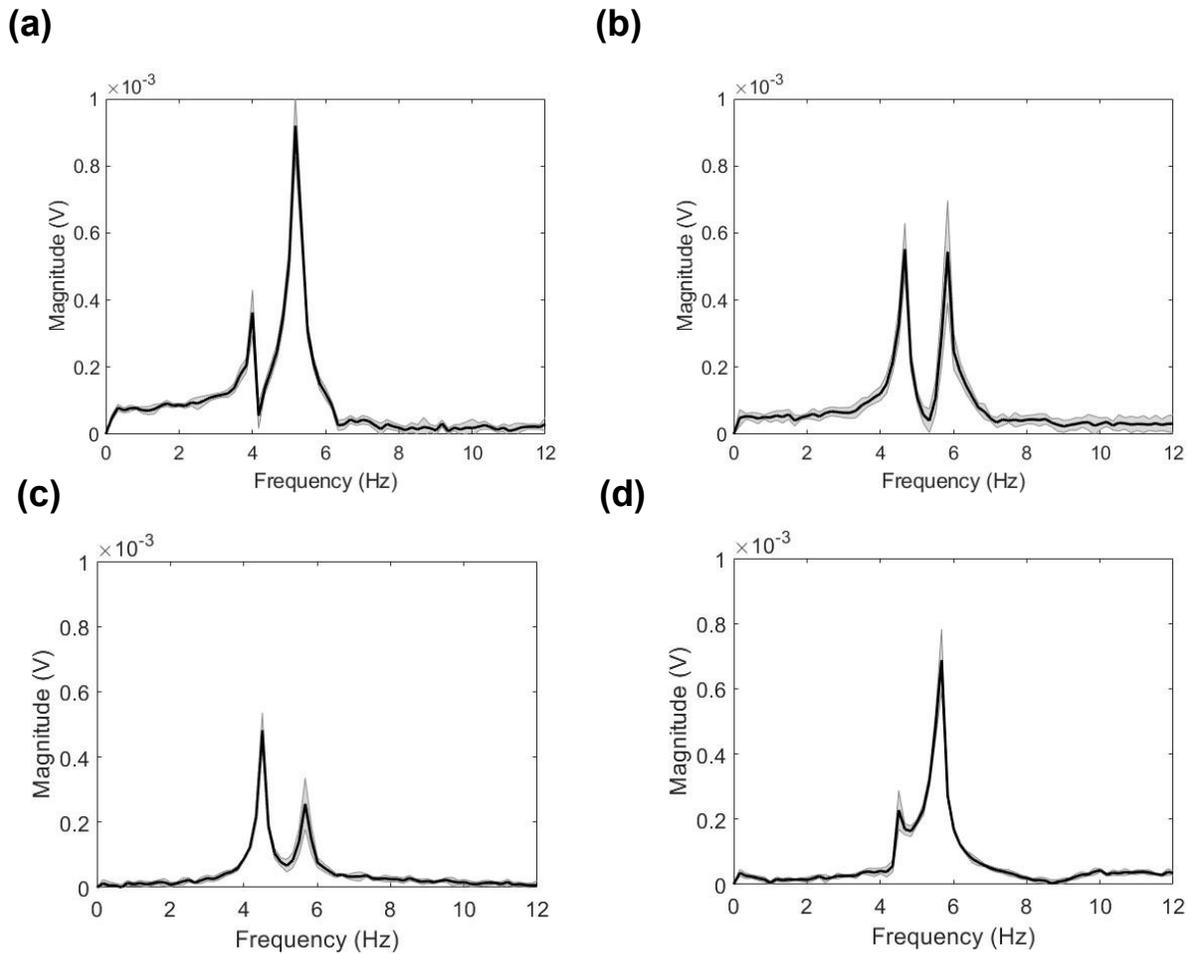

**Figure 14. A comparison of the same treatment, where the spider robot dynamically crouches on the web with a stationary prey robot on spiral thread 13 at different times.** The prey robot is placed at the same location on the web, but the response is different. However, the overall structure of the vibration profile remains the same, there are still two-peak frequencies. Each example is taken around 1–3 months apart. **(a)** Earliest example taken. Peak frequencies are at 4 and 5.2 Hz. **(b)** Sample taken ~3 months after Fig. 14(a). Peak frequencies are at 4.7 Hz and 5.8 Hz. **(c)** Sample taken ~1 month after Fig. 14(b). Peak frequencies are at 4.5 Hz and 5.7 Hz. **(d)** Sample used in Fig. 6(b). This was taken ~1 month after Fig. 14(c). Peak frequencies are at 4.5 Hz and 5.7 Hz.

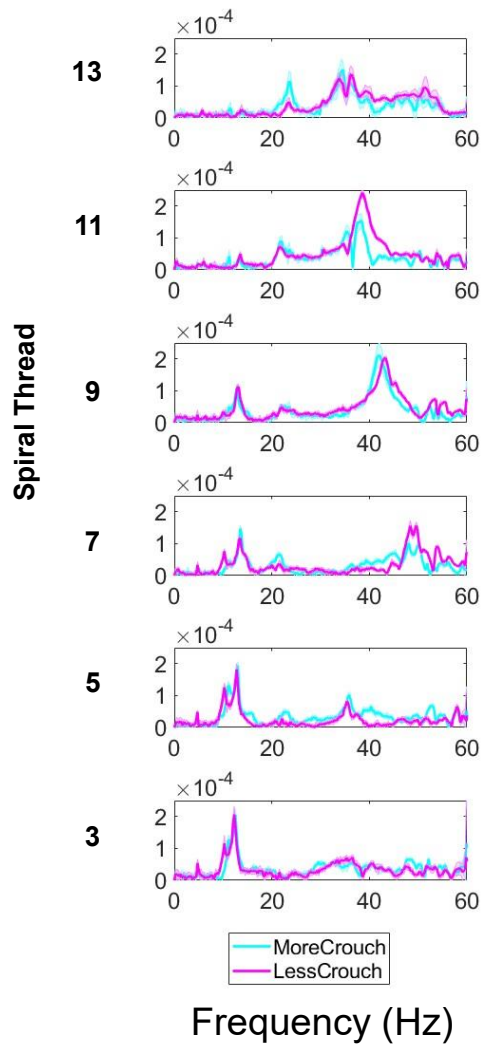

**Figure 15. FFTs of Accelerometer D2 when spider robot is not actively moving and prey robot was actively moving once at different spiral threads.** Being in a more crouched state or less crouched state did not seem to affect frequencies significantly. Compared to dynamic crouching, the location is much harder to tell using the vibration spectra sensed by the spider robot when the spider robot is static. However, there are still some identifiable peaks, including 2 ~12 Hz components and a ~22 Hz component that shift when the prey robot changes location, and a 5 Hz and >30Hz components that change randomly. As the prey robot moves closer to the center of the web, the ~22 Hz peak decreases in magnitude, letting the ~12 Hz components increase in magnitude instead.

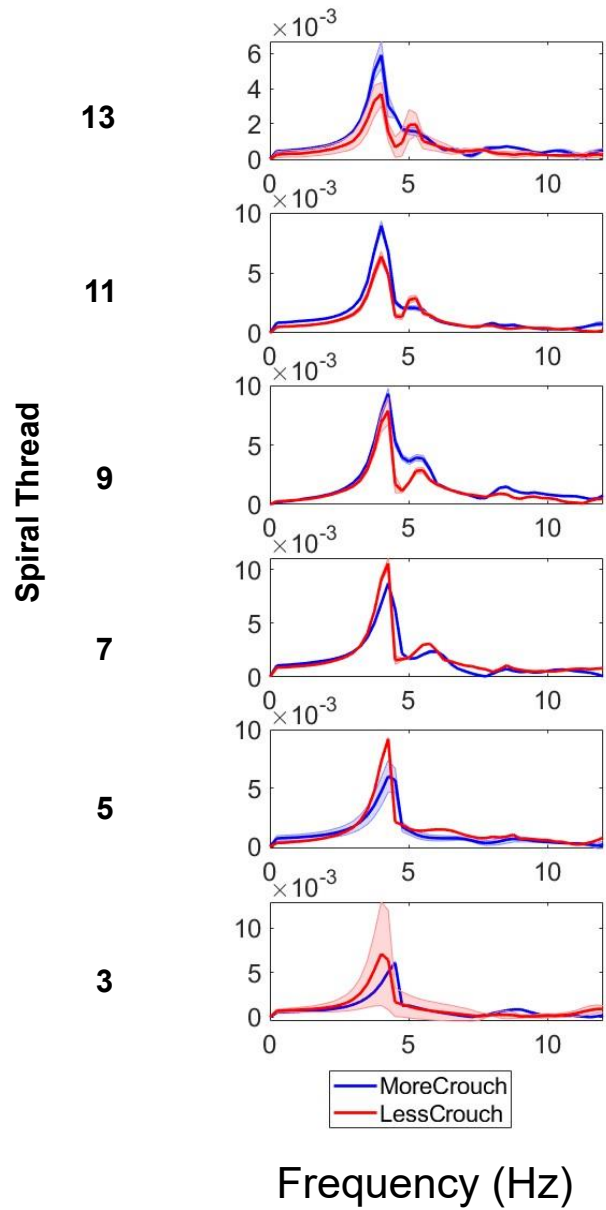

**Figure 16. FFTs of Accelerometer D2 when spider robot is not actively moving and prey robot impacts web hard on different spiral threads.** Second frequencies are visible in most treatments and reflect main experiment findings, in that the $2^{nd}$ peak frequency gets higher in frequency and lower in magnitude as prey robot shifts closer to center of web.

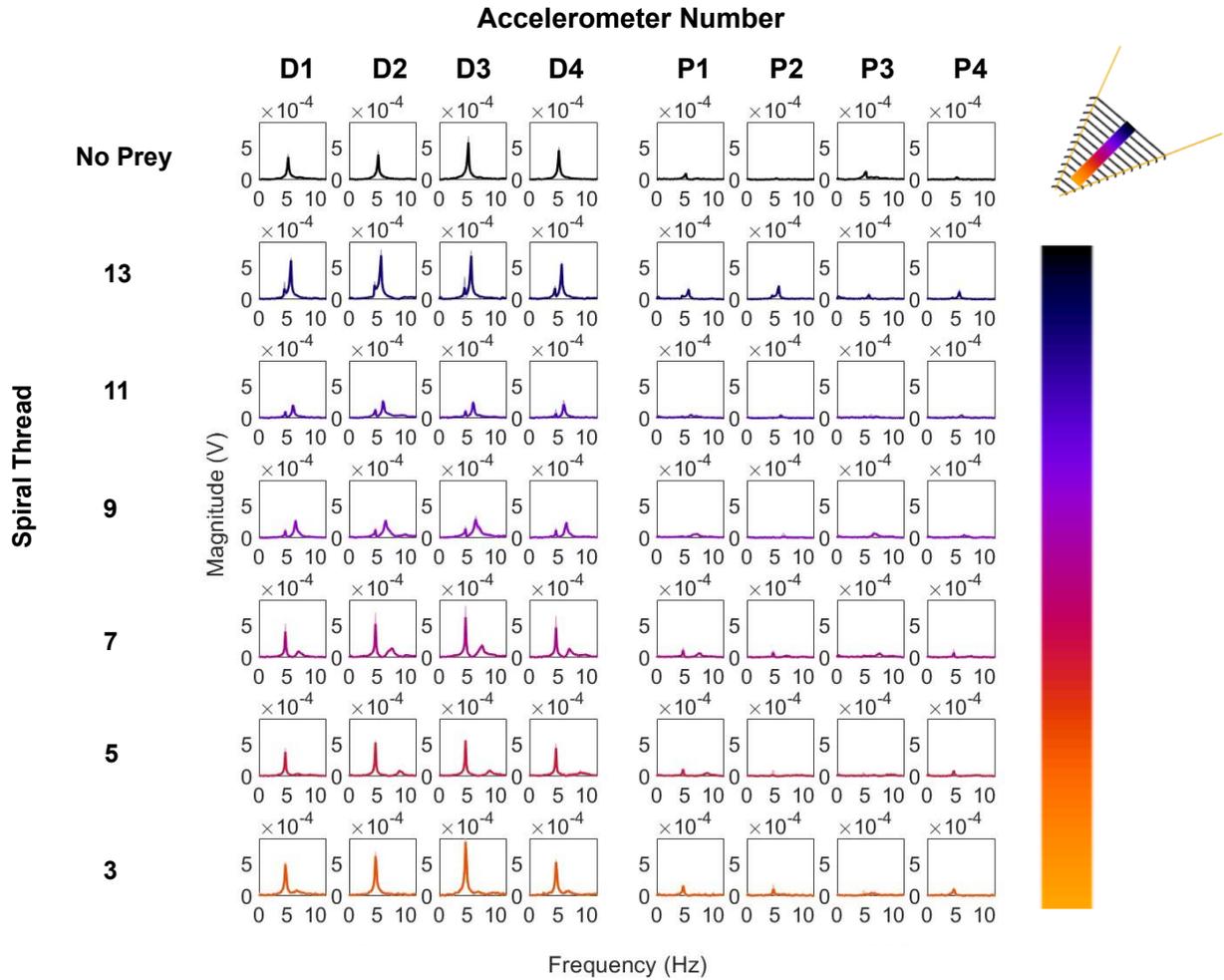

**Figure 17. All FFTs of all accelerometers when spider robot is jerky crouching once with a not actively moving prey robot.** All distal accelerometers roughly show the same two peak frequency. Some accelerometers in certain cases show a weaker second frequency. Proximal accelerometers have a weak response and thus are not mentioned in the paper.

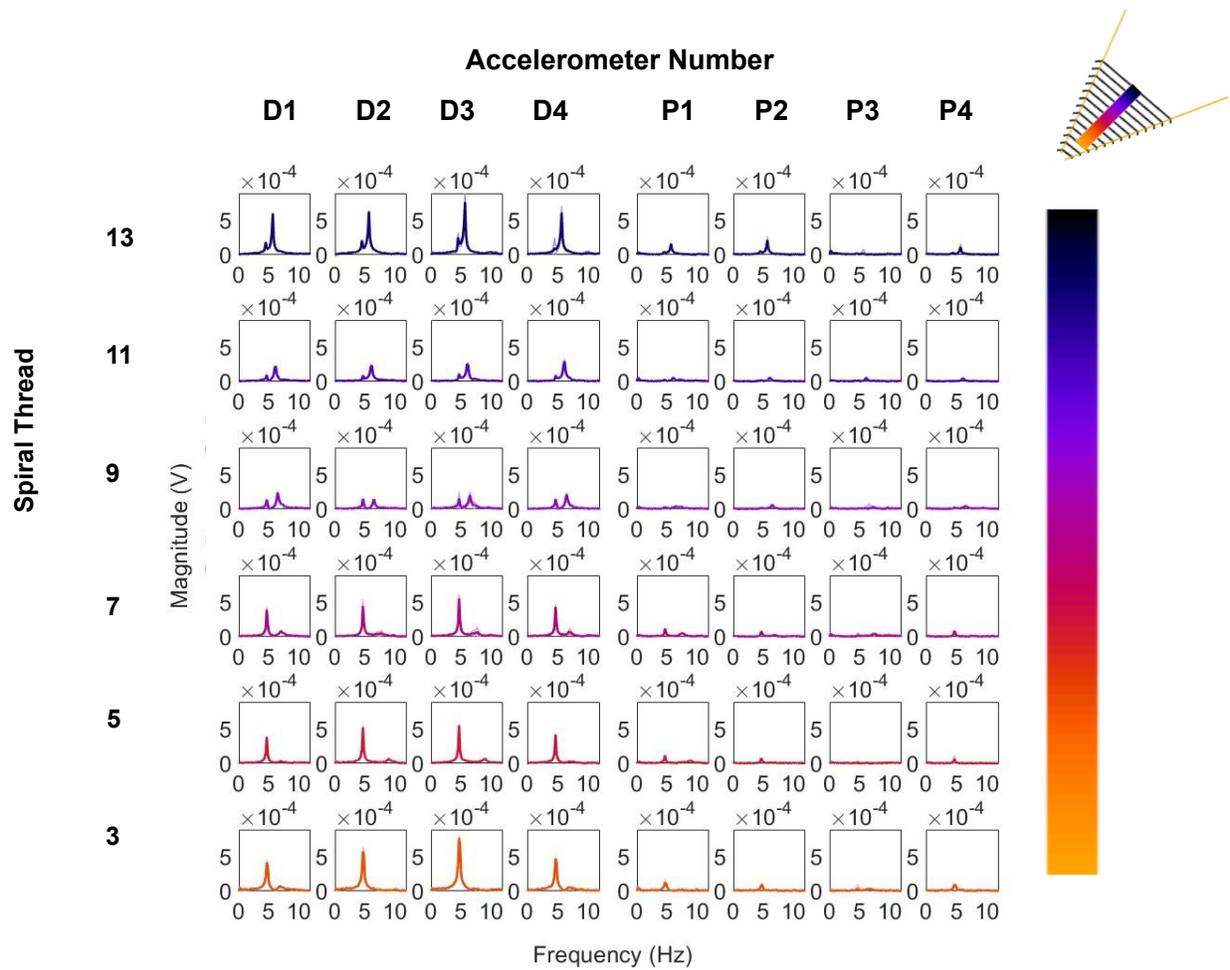

**Figure 18. All FFTs of all accelerometers when spider robot is jerky crouching once with a prey robot that actively shakes once**. All distal accelerometers roughly show the same two peak frequency. Some accelerometers in certain cases show a weaker second frequency. The behavior is very similar to Fig. 17.

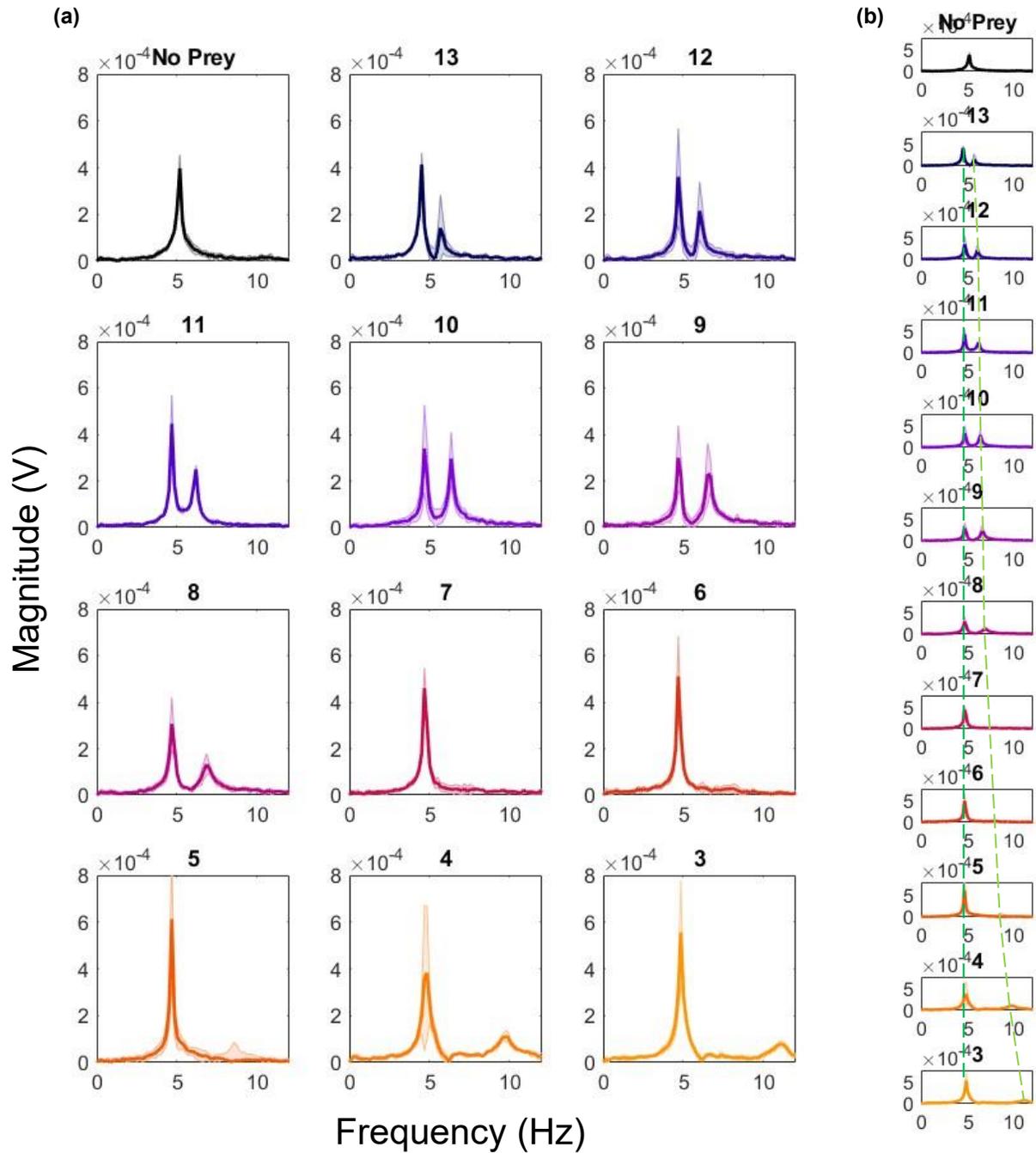

**Figure 19. Another experiment setup in the same conditions as Fig. 6(b) taken ~1 month earlier, with more spiral threads tested.** The two-peak frequency exists in most spiral threads, except for spiral threads 6 and 7. The cause for the consistent disappearance of the second frequency is unknown, but unlike Fig.

4B, the second frequency is much clearer. While the frequency values and the magnitude differ, the same trends in Fig. 4B can still be observed in this experiment's results. This demonstrates that our discovered mechanism is robust despite subtle environmental changes. This was not included in main experiment results as web vibrations were not analyzed.

**Table 1. Data used in Fig. 11(c) and (d) for correlation calculations.** The frequencies are captured from Fig. 6(b) and Fig. 19.

| Spiral thread | Length of spiral thread (m) | 1/sqrt(Length) | Frequency from main experiment Fig. 6(b) (Hz) | Frequencies from other experiment Fig. 19 (Hz) |
|---|---|---|---|---|
| 13 | 0.293 | 1.847 | 5.676 | 5.676 |
| 12 | 0.273 | 1.914 |  | 6.010 |
| 11 | 0.250 | 2.000 | 6.010 | 6.177 |
| 10 | 0.225 | 2.108 |  | 6.344 |
| 9 | 0.205 | 2.209 | 6.511 | 6.678 |
| 8 | 0.185 | 2.325 |  | 7.011 |
| 7 | 0.165 | 2.462 | 7.011 |  |
| 6 | 0.145 | 2.626 |  |  |
| 5 | 0.125 | 2.828 | 9.015 | 8.680 |
| 4 | 0.105 | 3.086 |  | 10.016 |
| 3 | 0.085 | 3.430 | 10.183 | 11.018 |
| 2 | 0.065 | 3.922 |  |  |
| 1 | 0.045 | 4.714 |  |  |

robotics, soft matter and dynamical systems, Reports Prog. Phys. **79**, 110001 (2016).

[67] N. Gravish and G. V. Lauder, Robotics-inspired biology, J. Exp. Biol. **221**, 1 (2018).

[68] *Physics of Life* (2023).

[69] J. Long, Darwin's Devices: What Evolving Robots can Teach Us about the History of Life and the Future of Technology. John Long., Integr. Comp. Biol. **52**, 546 (2012).

[70] A. J. Ijspeert, Biorobotics: Using robots to emulate and investigate agile locomotion, Science (80-. ). **346**, 196 (2014).

[71] B. Webb, What does robotics offer animal behaviour?, Anim. Behav. **60**, 545 (2000).

[72] B. Webb, Can robots make good models of biological behaviour?, Behav. Brain Sci. **24**, 1033 (2001).

[73] S. Sun, E. H. Lin, N. Brown, H.-Y. Hung, A. Gordus, J. Mueller, and C. Li, Creating a Biologically More Accurate Spider Robot to Study Active Sensing, 2025.

[74] W. G. Eberhard, The Ecology of the Web of Uloborus Diversus (Araneae: Uloboridae)*, Springer-Verlag, 1971.

[75] W. G. Eberhard, The Webs of Newly Emerged Uloborus Diversus and of a Male Uloborus sp . ( Araneae : Uloboridae ) Author ( s ): William G . Eberhard Published by : American Arachnological Society Stable URL : http://www.jstor.com/stable/3705018, **4**, 201 (1976).

[76] T. Hesselberg, Exploration behaviour and behavioural flexibility in orb-web spiders: A review, Curr. Zool. **61**, 313 (2015).

[77] W. G. Eberhard, Function and Phylogeny of Spider Webs, Annu. Rev. Ecol. Evol. Syst. **21**, 341 (1990).

[78] R. Hergenröder and F. G. Barth, Vibratory signals and spider behavior: How do the sensory inputs
82

bending., Bioinspir. Biomim. **18**, (2023).

[172] Movie S7. (Left) Prey robot (circled in red) vibrating passively at higher frequency after spider robot's jerky crouch, when prey robot is not actively moving, played in real time. (Right) Prey vibrating passively at higher frequency after spider's jerky crouch, played at 0.2× speed.